\documentclass[12pt]{article}
\usepackage{amsfonts}
\usepackage{amsmath}
\usepackage{amsthm}
\usepackage{graphicx}
\usepackage[numbers,sort&compress]{natbib}
\numberwithin{equation}{section}
\title{Dbar-approach to coupled nonlocal NLS equation and general nonlocal Reduction}
\date{}
\author{Xueru Wang and Junyi Zhu\thanks{jyzhu@zzu.edu.cn}\\
\leftline{\hspace{0.6 cm}{\small{\sl School of Mathematics and Statistics, Zhengzhou University, 100 Kexue Road,}}}\\
\leftline{\hspace{0.6 cm}{\small{\sl  Zhengzhou, Henan 450001, People's Republic of China}}}}

\begin{document}
\maketitle {}
\begin{abstract}
The coupled nonlocal NLS equation is studied by virtue of the $2\times2$ Dbar-problem. Two spectral transform matrices are introduced to define two associated Dbar-problems. The relations between the coupled nonlocal NLS potential and the solution of the Dbar-problem are constructed. The spatial transform method is extended to obtain the coupled nonlocal NLS equation and its conservation laws. The general nonlocal reduction of the coupled nonlocal NLS equation to the nonlocal NLS equation is discussed in detail. The explicit solutions are derived.

\vspace{0.2cm}


%%\vspace{0.2cm}
%Mathematics Subject Classification: 35B20
\end{abstract}
\section{Introduction}\vspace{0.3cm}
\hspace{0.5cm}
%A new integrable nonlocal nonlinear Schr\"{o}dinger (NLS) equation is introduced in \cite{prl110-064105}.
By virtue of a novel left-right Riemann-Hilbert problem, the inverse scattering transform of nonlocal NLS (nNLS) equation and a circumstantial comparison with the classical NLS equation are given in \cite{prl110-064105,non29-915}.
The inverse scattering transform for the nNLS equation with nonzero
boundary conditions at infinity is presented in \cite{jmp59-011501,non31-5385}.
Under the \textit{PT}-symmetric transformation, coupled nonlocal NLS equation and general vector nonlocal NLS equation are discussed in \cite{aml62-101,aml103-106209}. Alice-Bob systems are introduced in \cite{cpl34-100201,jmp59-083507}. The long-time behavior of the nonlocal NLS equation was considered in \cite{jmp60-031504}. In particular, the nNLS equation admits both bright and dark solitons \cite{pre89-052918}.
The higher-order rational solitons of the nNLS equation are given in \cite{chaos26-063123,aml69-113}.
Rogue waves in the nonlocal \textit{PT}-symmetric nonlinear Schrodinger equation are given in \cite{lmp109-945,jmaa487-124023}.
The multi-linear form and some self-similar solutions are investigated in \cite{aml47-61}.
Discrete nonlocal NLS equation was presented in  \cite{pre90-032912,pre89-052918,non33-3653,jmp57-083507}.
The reverse-time nNLS equations are discussed in \cite{sam139-7,pla383-328,aml102-106161,sam145-197,pams149-251,sam145-563}.
A nonlocal derivative nonlinear Schr\"odinger equation is introduced \cite{1612.04892}.
Transformations between nonlocal and local integrable equations are presented in \cite{sam140-178}.
Nonlocal reductions for nonlocal integrable equations are investigate in \cite{jmp59-051501,cnsns67-427,cnsns71-161}.

$\bar\partial$ (Dbar)-problem is an effective tool to study nonlinear evolution equations and to give their explicit solutions \cite{pd18-242,ip4-123,jpa21-L537,D-L2007,jpa46-035204,mpag17-49,aml66-47,ctp72-015004}.
Recently, we extended the Dbar approach to study the NLS equation with nonzero boundary condition \cite{2011.09028}.
The existing research methods to nonlocal integrable equations are mainly the inverse scattering method (the Riemann-Hilbert problem) and the Darboux transformation.
While, the Dbar-problem to investigate the nonlocal integrable equation is still an open problem.
In this paper, we extend the Dbar-approach to investigate the coupled nonlocal NLS (cnNLS) equation
%Nonlocal nonlinear Schr\"{o}dinger equation
\begin{equation}\label{a1}
\begin{aligned}
iq_{t}(x,t)&=q_{xx}(x,t)-2\sigma q^2(x,t)\overline{\hat{q}(-x,t)}, \\
i\hat{q}_{t}(x,t)&=\hat{q}_{xx}(x,t)-2\sigma \hat{q}^2(x,t)\overline{q(-x,t)}, \quad \sigma=\mp1.
\end{aligned}
\end{equation}
It is noted that, for the cnNLS equation (\ref{a1}), if $\{q(x,t),\hat{q}(x,t)\}$ is a set of solution so is $\{\overline{q(x,-t)},\overline{\hat{q}(x,-t)}\}$, and so is $\{q(-x,t),\hat{q}(-x,t)\}$.
In addition, if let $V(x,t)=-2\sigma q(x,t)\overline{\hat{q}(-x,t)}$ and $\hat{V}(x,t)=-2\sigma \hat{q}(x,t)\overline{q(-x,t)}$, then $\hat{V}(x,t)=\overline{V(-x,t)}$.
%which means that the resulting equation remains invariant under the joint transformation of
%$$x\to-x,\quad t\to-t, \quad \{q,\hat{q}\}\to\{{q}^*,\hat{q}^*\}.$$

Equation (\ref{a1}) reduces to the nNLS equation \cite{prl110-064105,non29-915}
\begin{equation}\label{a1a}
\begin{aligned}
iq_{t}(x,t)&=q_{xx}(x,t)-2\sigma q^2(x,t)\overline{{q}(-x,t)},
%i\hat{q}_{t}(x,t)&=\hat{q}_{xx}(x,t)-2\sigma \hat{q}^2(x,t)\overline{q(-x,t)}, \quad \sigma=\mp1.
\end{aligned}
\end{equation}
if $q(x,t)=\hat{q}(x,t)$. We note that the cnNLS equation (\ref{a1}) is derived from a $2\times2$ matrix linear problem, so it is different from the multi-component or vector ones \cite{aml62-101,aml103-106209}.

%If we introduce a new function $u(x,t)=\big(q(x,t)+\hat{q}(x,t)\big)/2$, and a self-induced potential $V(x,t)=-\sigma[q(x,t)\overline{\hat{q}(-x,t)}+\hat{q}(x,t)\overline{q(-x,t)}]$, equation (\ref{a1}) takes the form
%\begin{equation}\label{a1b}
%iu_t=u_{xx}+V(x,t)u(x,t).
%\end{equation}

It is known that the relation between the NLS potential and the solution of the Dbar problem is established by the Dbar dressing method, among which a spectral transform matrix is introduced. The explicit solution can be given by choosing the spectral transform matrix with certain scattering data, which are called the Dbar data.
While for cnNLS equation, we have to define two different spectral transform matrices $R(k;x,t)$ and $\hat{R}(k;x,t)$, which give two associated Dbar problems $\bar\partial \psi(k;x,t)=\psi(k;x,t)R(k;x,t)$ and $\bar\partial \hat\psi(k;x,t)=\hat\psi(k;x,t)\hat{R}(k;x,t)$. With the Dbar-approach to the cnNLS equation with $\sigma=-1$, we show a simple and clear picture about the recostruction of the cnNLS potential about the scattering data which is equivalent to the Dbar data given by %. obtain two representations of the nonlocal NLS potential about the Dbar data
$\{\lambda_l, d_l\}_{l=1}^{\tilde{N}}$ and $\{k_j, c_j\}_{j=1}^N$ in the spectral transform matrices $R(k;x,t)$ and $\hat{R}(k;x,t)$. For the first spectral transform matrix, we have one set of representations
\begin{equation}\label{a2}
\begin{aligned}
q(x,t)&=2i\sum^{\tilde{N}}_{l=1}\bar{d}_l\mathrm e^{-2i\theta(-\bar\lambda_l;x,t)}{\psi}_{11}(-\bar\lambda_l;x,t),\\
r(x,t)&=-2i\sum^{N}_{j=1}c_j\mathrm e^{2i\theta(x,k_j;x,t)}{\psi}_{22}(k_j;x,t),
\end{aligned}
\end{equation}
and for the second spectral transform matrix, we obtain another set of representations
\begin{equation}\label{a3}
\begin{aligned}
\hat{q}(x,t)&=-2i\displaystyle\sum^{N}_{j=1}\bar{c}_j\mathrm e^{-2i\theta(-\bar{k}_j;x,t)}\hat{\psi}_{11}(-\bar{k}_j;x,t),\\
\hat{r}(x,t)&=2i\displaystyle\sum^{\tilde{N}}_{l=1}{d}_l\mathrm e^{2i\theta({\lambda}_l;x,t)}\hat{\psi}_{22}({\lambda}_l;x,t),
\end{aligned}
\end{equation}
where $\theta(k;x,t)=kx-2k^2t $.
It is noted that the eigenfunctions admit the following symmetry conditions
\begin{equation}\label{a4}
 \begin{aligned}
\hat\psi_{11}(k;x,t)=\overline{\psi_{22}(-\bar{k};-x,t)}, \quad  \hat\psi_{12}(k;x,t)=-\sigma\overline{\psi_{21}(-\bar{k};-x,t)},\\
\hat\psi_{21}(k;x,t)=-\sigma\overline{\psi_{12}(-\bar{k};-x,t)}, \quad \hat\psi_{22}(k;x,t)=\overline{\psi_{11}(-\bar{k};-x,t)}.
 \end{aligned}
\end{equation}
Then we find $r(x,t)=\sigma\overline{\hat{q}(-x,t)}$ and $\hat{r}(x,t)=\sigma\overline{{q}(-x,t)}$.
%Since the Dbar data are different ($\Im\lambda_j=-\Im{k_j}$), then the representations of $q$ in (\ref{a2}) and (\ref{a3}) give two different solution of nonlocal focusing NLS equation. As an example, the explicit solution of nonlocal NLS equation from $q$ in (\ref{a3}) is discussed by eliminating the eigenfunction $\hat\psi_{11}(-\bar{k}_j;x,t)$. In the elimination, the two sets of the Dbar data are encapsulated in the solution. We note that the two sets of Dbar data give another explicit solution from $q$ in (\ref{a2}), but they play different roles in each of the representations of the nonlocal NLS potential.
In addition, we extended the spatial transform method \cite{pla117-62} to find the cnNLS equation and its conservation laws.

It is remarked that the choice of the parameters $\{\lambda_l, d_l\}_{l=1}^{\tilde{N}}$ and $\{k_j, c_j\}_{j=1}^N$ for obtaining the explicit solutions of the cnNLS equation (\ref{a1}) is more free. While to construct the solution of nNLS equation (\ref{a1a}), one needs to consider the reduction and to introduce some constraint conditions on the parameters to make sure that $\hat{q}(x,t)=q(x,t)$. We note that the current nonlocal reductions are usually to construct the first few solutions ($N=1,2,3$) for nonlocal equation, but very few investigations for the general nonlocal reduction are presented. Here, we express the solution with two sets of special determinants of symmetry matrices and give a full discussion of the general nonlocal reduction for cnNLS equation. We show that the constraint conditions are $N=\tilde{N}$,  $k_j=ib_j, \lambda_j=i\eta_j$ are imaginary numbers, and $\prod\limits_{\tiny1\leq m<m'\leq N}(\eta_{m}-\eta_{m'})^2=\prod\limits_{\tiny1\leq l<l'\leq N}(b_{l'}-b_{l})^2$, as well as
\begin{equation}\label{a5}
 |c_j|^2=\frac{\prod\limits_{l=1}^N(\eta_l-b_j)^2}{\prod\limits_{s=1,s\neq j}^N(b_s-b_j)^2}, \quad
 |d_j|^2=\frac{\prod\limits_{l=1}^N(\eta_j-b_l)^2}{\prod\limits_{s=1,s\neq j}^N(\eta_s-\eta_j)^2}.
\end{equation}
%and
%\begin{equation}\label{a6}
%\prod\limits_{\tiny1\leq m<m'\leq N}(\eta_{m}-\eta_{m'})^2=\prod\limits_{\tiny1\leq l<l'\leq N}(b_{l'}-b_{l})^2.
%\end{equation}
%$|c_j|,|d_j|$ are certain functions of $b_j$ and $\eta_j$, $(b_j<0<\eta_j)$.

The outline of this paper is as follows. In section 2, we introduce two local Dbar problems.
In section 3, we derive the focusing/defocusing cnNLS equation and its conservation laws.
In section 4, we present the explicit solutions for the focusing cnNLS equation. In section 5,
we discuss the nonlocal reductions to the nNLS equation in detail.

\section{Double Dbar-problems}\vspace{0.3cm}
\hspace{0.5cm}
Consider the first local Dbar-problem
\begin{equation}\label{c1}
\begin{aligned}
\bar{\partial}\psi(k)=\psi(k)R(k),
\end{aligned}
\end{equation}
with the normalization condition
\begin{equation}\label{c2}
\begin{aligned}
\psi(k)\rightarrow I,\quad k\rightarrow\infty,
\end{aligned}
\end{equation}
where $R(k)$ is the spectral transform matrix.
The Dbar-problem (\ref{c1}) and (\ref{c2}) equivalent to the following integral equation
\begin{equation}\label{c3}
\begin{aligned}
\psi(k)=I+\psi(k)R(k)C_k,\\
\end{aligned}
\end{equation}
where the Cauchy-Green operate in complex plane is defined as
\begin{equation}\label{c4}
\begin{aligned}
\psi(k)R(k)C_k=\frac{1}{2\pi i} \int\int\frac{dz\wedge d\bar{z}}{z-k}\psi(k)R(k).\\
\end{aligned}
\end{equation}

The aim of dressing method is construct the relation between the cnNLS potential and the solution of the Dbar-problem. To this end, a good way is to construct the cnNLS equation and its Lax pair from the Dbar-problem.
It is noted that the Dbar-problem is defined in the spectral space, while the cnNLS equation is in the physical space. Thus we need to introduce the physical variables $x,t$ into the function $\psi(k)$, which can be done by extending the spectral transform matrix to be the form $R(k;x,t)$, and letting %that it satisfies the following evolution equations
\begin{equation}\label{c5}
\begin{aligned}
R_x(k;x,t)=-ik[\sigma_3,R(k;x,t)],
\end{aligned}
\end{equation}
%and
\begin{equation}\label{c5a}
\begin{aligned}
R_t(k;x,t)=2ik^2[\sigma_3,R(k;x,t)].
%&\psi_t(x,t,k)=2ik^2[\sigma_3,\psi(x,t,k)]-2kQ(x)\psi(x,t,k)+i\sigma_3(Q^2(x)-Q_x(x))\psi(x,t,k).\\
\end{aligned}
\end{equation}
We note that the solution of the system (\ref{c5}) and (\ref{c5a}) is not unique.

Under the dressing procedure \cite{D-L2007, jpa46-035204,mpag17-49}, we find that
\begin{equation}\label{c6}
\begin{aligned}
&\psi_x(k;x,t)=-ik[\sigma_3,\psi(k;x,t)]+Q(x,t)\psi(k;x,t),\\
&Q(x,t)=-i[\sigma_3,\langle\psi(k;x,t) R(k;x,t)\rangle],\\
\end{aligned}
\end{equation}
and
\begin{equation}\label{c15}
\begin{aligned}
\psi_t(k;x,t)=&2ik^2[\sigma_3,\psi(k;x,t)]-2kQ(x,t)\psi(k;x,t)\\
&+i\sigma_3[Q^2(x,t)-Q_x(x,t)]\psi(k;x,t).
\end{aligned}
\end{equation}
where
\begin{equation}\label{c7}
\begin{aligned}
\langle\psi(k;x,t) R(k;x,t)\rangle=\frac{1}{2\pi i} \int\int\psi(k;x,t)R(k;x,t)dk\wedge d\bar{k}.\\
\end{aligned}
\end{equation}

For the cnNLS equation, we need to consider the second local Dbar problem
\begin{equation}\label{c8}
\begin{aligned}
&\bar{\partial}\hat{\psi}(k;x,t)=\hat{\psi}(k;x,t)\hat{R}(k;x,t),\\
&\hat{\psi}(k;x,t)\rightarrow I,\quad k\rightarrow\infty,
\end{aligned}
\end{equation}
where the new spectral transform matrix $\hat{R}(k;x,t)$ is another solution of the
evolution system (\ref{c5}) and (\ref{c5a}). Then we have
\begin{equation}\label{c9}
\begin{aligned}
\hat{\psi}(k;x,t)=I+\hat{\psi}(k;x,t)\hat{R}(k;x,t)C_k.\\
\end{aligned}
\end{equation}
A similar procedure gives another potential $\hat{Q}(x,t)$
\begin{equation}\label{c11}
\begin{aligned}
%&\hat{\psi}_x(k;x,t)=-ik[\sigma_3,\hat{R}(k;x,t)]+\hat{Q}(x)\hat{\psi}(k;x,t),\\
&\hat{Q}(x,t)=-i[\sigma_3,\langle\hat{\psi}(k;x,t)\hat{R}(k;x,t)\rangle],
\end{aligned}
\end{equation}
and the another linear spectral system
\begin{equation}\label{c13}
\begin{aligned}
&\hat{\psi}_x(k;x,t)=-ik[\sigma_3,\hat{\psi}(k;x,t)]+\hat{Q}(x,t)\hat{\psi}(k;x,t),\\
%&\hat{Q}(x)=-ik[\sigma_3,\langle\hat{\psi}(k;x,t)\hat{R}(k;x,t)\rangle].\\
\end{aligned}
\end{equation}
and
\begin{equation}\label{c16}
\begin{aligned}
\hat\psi_t(k;x,t)=&2ik^2[\sigma_3,\hat\psi(k;x,t)]-2k\hat{Q}(x,t)\hat\psi(k;x,t)\\
&+i\sigma_3[\hat{Q}^2(x,t)-\hat{Q}_x(x,t)]\hat\psi(k;x,t).
\end{aligned}
\end{equation}

In addition, to get the cnNLS equation, one also needs to introduce a symmetry condition about the two potentials
\begin{equation}\label{c17}
 \hat{Q}(x,t)=-\Lambda\overline{Q(-x,t)}\Lambda^{-1}, \quad \Lambda=\left(\begin{matrix}
 0&1\\
 -\sigma&0
 \end{matrix}\right),
\end{equation}
then the symmetry condition of the eigenfunction takes the following form
\begin{equation}\label{c18}
 \hat\psi(k;x,t)=\Lambda\overline{\psi(-\bar{k};-x,t)}\Lambda^{-1}.
\end{equation}
Thus, we have
\begin{equation}\label{c19}
 Q=\left(\begin{matrix}
 0&q(x,t)\\
 \sigma\overline{\hat{q}(-x,t)}&0
 \end{matrix}\right),\quad \hat{Q}=\left(\begin{matrix}
 0&\hat{q}(x,t)\\
 \sigma\overline{{q}(-x,t)}&0
 \end{matrix}\right),
\end{equation}
and %$\langle\hat\psi(k;x,t)\hat{R}(k;x,t)\rangle_{ij}=\langle\psi(k;x,t){R}(k;x,t)\rangle_{jl}, (j\neq l)$ as well as
\begin{equation}\label{c20}
 \begin{aligned}
\hat\psi_{11}(k;x,t)=\overline{\psi_{22}(-\bar{k};-x,t)}, \quad  \hat\psi_{12}(k;x,t)=-\sigma\overline{\psi_{21}(-\bar{k};-x,t)},\\
\hat\psi_{21}(k;x,t)=-\sigma\overline{\psi_{12}(-\bar{k};-x,t)}, \quad \hat\psi_{22}(k;x,t)=\overline{\psi_{11}(-\bar{k};-x,t)}.
 \end{aligned}
\end{equation}

\section{nNLS equation and conservation laws}\vspace{0.3cm}
\hspace{0.5cm}
From (\ref{c2}), we know that $\psi(k;x,t)$ has the following asymptotic behaviors
\begin{equation}\label{g1}
\begin{aligned}
\psi(k;x,t)=\displaystyle\sum^{\infty}_{n=0}\frac{a_n(x,t)}{k^n},\quad k\rightarrow\infty,\\
\end{aligned}
\end{equation}
where $a_0(x,t)=I.$
In view of the temporal linear spectral problem (\ref{c15}), we get
\begin{equation}\label{g3}
\begin{aligned}
\psi_t(k;x,t)+2ik^2\psi(k;x,t)\sigma_3=T(k;x,t)\psi(k;x,t),\\
\end{aligned}
\end{equation}
where
\begin{equation}\label{g4}
\begin{aligned}
T(k;x,t)=2ik^2\sigma_3-2kQ(x)+i\sigma_3(Q^2(x)-Q_x(x)).\\
\end{aligned}
\end{equation}

Let the trace of the spectral transform matrix $R(k;x,t)$ is zero, then the Dbar problem (\ref{c1}) implies
$
\bar{\partial}\det\psi(k;x,t)=0,
$
and further
$
\det\psi(k;x,t)=1
$
in view of the asymptotic behaviors (\ref{g1}).
Since $\psi^{-1}=\sigma_2\psi^T\sigma_2$, equation (\ref{g3}) can be rewritten as
\begin{equation}\label{g7}
\begin{aligned}
T(k;x,t)=\psi_t(k;x,t)\sigma_2\psi^T(k;x,t)\sigma_2+2k^2\Psi(k;x,t)\sigma_2,\\
\end{aligned}
\end{equation}
where
\begin{equation}\label{g8}
\begin{aligned}
\Psi=\psi\sigma_1\psi^T.\\
\end{aligned}
\end{equation}
It is noted that
\begin{equation}\label{g9}
\begin{aligned}
\Psi(k;x,t)=\displaystyle\sum^{\infty}_{n=0}\frac{\Psi_n(x,t)}{k^n},\quad k\rightarrow\infty,\\
\end{aligned}
\end{equation}
where
\begin{equation}\label{g10}
\begin{aligned}
\Psi_0(x,t)=\sigma_1,\quad \Psi_n(x,t)=\displaystyle\sum^{n}_{m=0}a_m\sigma_1a^T_{n-m},\quad (n\geq1).\\
\end{aligned}
\end{equation}

Substituting the expansions (\ref{g1}) and (\ref{g9}) into (\ref{g7}), and considering the $O(k^{-1})$ items, we obtain
\begin{equation}\label{g11}
\begin{aligned}
a_{1,t}(x,t)=-2\Psi_3(x,t)\sigma_2.\\
\end{aligned}
\end{equation}
In view of the spatial linear spectral problem (\ref{c6}), $\Psi$ satisfies
\begin{equation}\label{g12}
\begin{aligned}
\Psi_x(k;x,t)=-ik[\sigma_3\Psi(k;x,t)+\Psi(k;x,t)\sigma_3]+Q\Psi(k;x,t)+\Psi(k;x,t)Q^T,\\
\end{aligned}
\end{equation}
which can be rewritten as
\begin{equation}\label{g13}
\begin{aligned}
\Psi_x^{[d]}(k;x,t)=-2ik\sigma_3\Psi^{[d]}(k;x,t)+Q\Psi^{[o]}(k;x,t)+\Psi^{[o]}(k;x,t)Q^T,\\
\end{aligned}
\end{equation}
and
\begin{equation}\label{g14}
\begin{aligned}
\Psi_x^{[o]}(k;x,t)=Q\Psi^{[d]}(k;x,t)+\Psi^{[d]}(k;x,t)Q^T,\\
\end{aligned}
\end{equation}
where $\Psi^{[d]}$ and $\Psi^{[o]}$ denote the diagonal part and off-diagonal part of the matrix $\Psi$.
Substituting the expansion (\ref{g9}) into (\ref{g13}) and (\ref{g14})
\begin{equation}\label{g15}
\begin{aligned}
&\Psi^{[d]}_1=-i\sigma_3Q\sigma_1,\\
&\Psi^{[d]}_{n+1}=\frac{i}{2}\sigma_3\Psi^{[d]}_{n,x}-\frac{i}{2}\sigma_3\big(Q\Psi_n^{[o]}+\Psi_n^{[o]}Q^T\big),\\
&\Psi^{[o]}_{n,x}=\big(Q\Psi_n^{[d]}+\Psi_n^{[d]}Q^T\big).\\
\end{aligned}
\end{equation}
For simplicity, here and after we omit the variables $(x,t)$. From the above recurrent formula, we find
\begin{equation}\label{g16}
\begin{aligned}
&\Psi^{[d]}_1=-Q\sigma_2,\quad \Psi^{[o]}_1=0,\\
&\Psi^{[d]}_2=\frac{1}{2}Q_x\sigma_1,\quad \Psi^{[o]}_2=\frac{1}{2}Q^2\sigma_1,\\
&\Psi^{[d]}_3=\frac{1}{4}Q_{xx}\sigma_2-\frac{1}{2}Q^2Q\sigma_2,\quad \Psi^{[o]}_3=\frac{1}{4}(QQ_x-Q_xQ)\sigma_2,\\
&\Psi^{[d]}_4=(-\frac{1}{8}Q_{xxx}+\frac{3}{4}Q^2Q_x)\sigma_1,\\
&\Psi^{[o]}_4=-\frac{1}{8}(QQ_{xx}+Q_{xx}Q-Q^2_x-3Q^4)\sigma_1.\\
\end{aligned}
\end{equation}

From the off-diagonal part of equation (\ref{g11}), we obtain the nonlinear equation
\begin{equation}\label{g17}
\begin{aligned}
i\sigma_3Q_t-Q_{xx}+2Q^2Q=0,\\
\end{aligned}
\end{equation}
which implies the nonlocal nonlinear Schr\"{o}dinger equation (\ref{a1}). From the diagonal part of (\ref{g11}), we find the first conservation law
\begin{equation}\label{g18}
\begin{aligned}
i(qr)_t=(q_xr-qr_x)_x, \quad r=\sigma\overline{\hat{q}(-x,t)}.
%i(\hat{q}\hat{r})_t=(\hat{q}_x\hat{r}-\hat{q}\hat{r}_x)_x, \quad \hat{r}=\sigma\overline{{q}(-x,t)}.
\end{aligned}
\end{equation}

Similarly, the $O(k^{-2})$ terms in the expansion of (\ref{g7}) have the following form
\begin{equation}\label{g19}
\begin{aligned}
a_{2,t}+a_{1,t}\sigma_2a_1^T\sigma_2+2\Psi_4\sigma_2=0,\\
\end{aligned}
\end{equation}
The off-diagonal part of equation (\ref{g19}) also implies the equation (\ref{g17}), and the diagonal part gives the second conservation law
\begin{equation}\label{g20}
\begin{aligned}
i(qr_x)_t=(q_xr_x+q^2r^2-qr_{xx})_x, \quad r=\sigma\overline{\hat{q}(-x,t)}.
\end{aligned}
\end{equation}
%The following results have been used
%\begin{equation}\label{g21}
%\begin{aligned}
%&a_1^{[0]}=-\frac{i}{2}\sigma_3Q,\quad
%a_{1,x}^{[d]}=\frac{i}{2}\sigma_3Q^2,\\
%&a_2^{[0]}=\frac{1}{4}(Q_x-Q\partial^{-1}_xQ^2),\quad
%a_{2,x}^{[d]}=\frac{1}{4}(QQ_x-Q^2\partial^{-1}_xQ^2).\\
%\end{aligned}
%\end{equation}
The more conversation laws of the cnNLS equation can be derived similarly from (\ref{g7}) \cite{pla117-62}.
The same results can be derived from the second linear system (\ref{c13}) and (\ref{c16}).

\section{The solutions of coupled nonlocal focusing NLS equation}\vspace{0.3cm}
\hspace{0.5cm}
In this section, we give the explicit solutions of cnNLS equation in the case $\sigma=-1$.

According to the above symmetry conditions, we let the first spectral transform matrix ${R}(k;x,t)$ has the following form
\begin{equation}\label{d1}
\begin{aligned}
&&
{R}(k;x,t)=\pi\left(\begin{array}{cc}
0&\displaystyle\sum^{\tilde{N}}_{j=1}\bar{d}_j\mathrm e^{-2i\theta(k;x,t)}\delta(k+\bar\lambda_j)\\
\displaystyle\sum^{N}_{j=1}c_j\mathrm e^{2i\theta(k;x,t)}\delta(k-k_j)&0\\
\end{array}\right),
\end{aligned}
\end{equation}
and take the second spectral transform matrix be of the form
\begin{equation}\label{d2}
\begin{aligned}
&&
{\hat{R}}(k;x,t)=-\pi
\left(\begin{array}{cc}
0&\displaystyle\sum^{N}_{j=1}\bar{c}_j\mathrm e^{-2i\theta(k;x,t)}\delta(k+\bar{k}_j)\\
\displaystyle\sum^{\tilde{N}}_{j=1}{d}_j\mathrm e^{2i\theta(k;x,t)}\delta(k-{\lambda}_j)&0\\
\end{array}\right),
\end{aligned}
\end{equation}

%\begin{equation}\label{}
%\begin{aligned}
%&&
%{\hat{R}}(k;x,t)=-\pi
%\left(\begin{array}{cc}
%0&\displaystyle\sum^{N}_{j=1}\bar{c}_j\mathrm e^{-2i\theta(k;x,t)}\delta(k-{k}_j)\\
%\displaystyle\sum^{\tilde{N}}_{j=1}{d}_j\mathrm e^{2i\theta(k;x,t)}\delta(k+{\bar{\lambda}}_j)&0\\
%\end{array}\right),
%\end{aligned}
%\end{equation}

where
\begin{equation}\label{d3}
\theta(k;x,t)=kx-2k^2t.
\end{equation}

From equations (\ref{d1}) and (\ref{c6}), we get
\begin{equation}\label{d4}
\begin{aligned}
q(x,t)&=2i\sum^{\tilde{N}}_{l=1}\bar{d}_l\mathrm e^{-2i\theta(-\bar\lambda_l;x,t)}{\psi}_{11}(-\bar\lambda_l;x,t),\\
r(x,t)&=-2i\sum^{N}_{j=1}c_j\mathrm e^{2i\theta(x,k_j;x,t)}{\psi}_{22}(k_j;x,t).
\end{aligned}
\end{equation}
Equations (\ref{d2}) and (\ref{c11}) imply another representations of the solution of the cnNLS equation
\begin{equation}\label{d5}
\begin{aligned}
\hat{q}(x,t)&=-2i\displaystyle\sum^{N}_{j=1}\bar{c}_j\mathrm e^{-2i\theta(-\bar{k}_j;x,t)}\hat{\psi}_{11}(-\bar{k}_j;x,t),\\
\hat{r}(x,t)&=2i\displaystyle\sum^{\tilde{N}}_{l=1}{d}_l\mathrm e^{2i\theta({\lambda}_l;x,t)}\hat{\psi}_{22}({\lambda}_l;x,t).
\end{aligned}
\end{equation}
Using the symmetry condition (\ref{c20}) and $\overline{\theta(-\bar{k};-x,t)}=\theta(k;x,t)$, we find that
%$q(x,t)$ in (\ref{d5}) and $r(x,t)$ in (\ref{d4}) admit
$r(x,t)=-\overline{\hat{q}(-x,t)}$ and $\hat{r}(x,t)=-\overline{{q}(-x,t)}$.
%another pair of $q(x,t)$ in (\ref{d4}) and $r(x,t)$ in (\ref{d5}) satisfy the same result.

%We note that the nNLS potential $q(x,t)$ is dependent on the scattering data, and, in general, different set of scattering data will give different solution. In this procedure, we give two sets of scattering data $\{\lambda_j, d_j\}$ and $\{{k}_j,{c}_j\}$, which imply two different solutions. It is noted that each solution will dependent on the two sets of scattering data at the same time.
%But the two sets of scattering data play the different role in different solution.
%In this construction, the two sets of scattering data are mixed in the integral equation (\ref{c9}) for $q(x,t)$ in (\ref{d5}), and combined in the integral equation (\ref{c3}) for $q(x,t)$ in (\ref{d4}).
%As an example, we give the explicit solution from the expression of $q(x,t)$ in (\ref{d5}).

Substituting (\ref{d1})  into (\ref{c3}), we obtain
\begin{equation}\label{d6a}
%\begin{aligned}
\psi(k;x,t)=I+\left(\sum^{N}_{j=1}\frac{g_j}{k-k_j}\psi^{[2]}(k_j;x,t),
\sum^{\tilde{N}}_{l=1}\frac{\hat{h}_l}{k+\bar{\lambda}_l}\psi^{[1]}(-\bar{\lambda}_l;x,t)\right),\\
\end{equation}
where $[j]$ denote the $j$-th column of $\psi(k;x,t)$.
Similarly, from (\ref{d2}) and (\ref{c9}), we have
\begin{equation}\label{d6b}
\hat\psi(k;x,t)=I-\left(\sum^{\tilde{N}}_{l=1}\frac{{h}_l}{k-\lambda_l}\hat{\psi}^{[2]}(\lambda_l;x,t),
\sum^{{N}}_{j=1}\frac{\hat{g}_j}{k+\bar{k}_j}\hat{\psi}^{[1]}(-\bar{k}_j;x,t)\right).
%\hat\psi(k;x,t)=I-(\displaystyle\sum^{\tilde{N}}_{j=1}\frac{{d}_je^{2i\theta(x,-\bar{\lambda}_j;x,t)x}}{k+\bar{\lambda}_j}\hat{\psi}^{[2]}(-\bar{\lambda}_j;x,t),\displaystyle\sum^{{N}}_{j=1}\frac{\bar{c}_je^{-2i\theta(x,{k}_j;x,t)x}}{k-{k}_j}\hat{\psi}^{[1]}({k}_j;x,t)),\\
%\end{aligned}
\end{equation}
Here we have used the following notations
\begin{equation}\label{e1}
 \begin{aligned}
 {g}_j={c}_j\mathrm e^{2i\theta({k}_j;x,t)},\quad \hat{g}_j=\bar{c}_j\mathrm e^{-2i\theta(-\bar{k}_j;x,t)},\\
 {h}_l={d}_l\mathrm e^{2i\theta({\lambda}_l;x,t)}, \quad \hat{h}_l=\bar{d}_l\mathrm e^{-2i\theta(-\bar{\lambda}_l;x,t)}.
 \end{aligned}
\end{equation}
We note that $\hat{g}_j(x)=\overline{g_j(-x)}, \hat{h}_l(x)=\overline{h_l(-x)}$, and equations (\ref{d6a}) and (\ref{d6b}) are equivalent to each other in views of the symmetry condition (\ref{c18}).
%Then using (\ref{c18}) the symmetry condition about $\psi(k;x,t), \hat\psi(k;x,t)$ , we find  $k_j,~\lambda_j$ are pure imaginary number.

Since $\hat\psi_{11}(-\bar{k}_j;x,t)$ can be obtained from (\ref{d6b}), and $\psi_{11}(-\bar{\lambda}_l;x,t)$ from (\ref{d6a}), %which can be done by substituting (\ref{d2}) into the integral equation (\ref{c9}). %In fact, we have
%\begin{equation}\label{d6}
%\begin{aligned}
%&\hat{\psi}(k;x,t)=I\\
%&+\sum^{N}_{j=1}
%\left(\begin{matrix}
%\frac{\bar{d}_j}{-\bar{\lambda}_j-k}\mathrm e^{2i\theta(-\bar{\lambda}_j;x,t)}\hat{\psi}_{12}(-\bar{\lambda}_j;x,t)&\frac{\bar{c}_j}{-\bar{k}_j-k}\mathrm e^{-2i\theta(-\bar{k}_j;x,t)}\hat{\psi}_{11}(-\bar{k}_j;x,t)\\
%\frac{\bar{d}_j}{-\bar{\lambda}_j-k}\mathrm e^{2i\theta(-\bar{\lambda}_j;x,t)}\hat{\psi}_{22}(-\bar{\lambda}_j;x,t)&\frac{\bar{c}_j}{-\bar{k}_j-k}\mathrm e^{-2i\theta(-\bar{k}_j;x,t)}\hat{\psi}_{21}(-\bar{k}_j;x,t)\\
%\end{matrix}\right).\\
%\end{aligned}
%\end{equation}
%From the (11) and (12) elements of the matrix equation (\ref{d6}) we can obtain the representation of $\hat\psi_{11}(-\bar{k}_j;x,t)$,
we find explicit solution for the cnNLS equation (\ref{a1})
\begin{equation}\label{e2}
q(x,t)=-2i\frac{\det \tilde{\Omega}^a}{\det{ \tilde\Omega}}, \quad \hat{q}(x,t)=2i\frac{\det {{\Omega}}^a}{\det{ {\Omega}}},
\end{equation}
where the symmetry matrix $\tilde\Omega$ and $\Omega$ are given by
\begin{equation}\label{e3}
\begin{aligned}
&\tilde\Omega_{\tilde{N}\times \tilde{N}}=\hat{H}^{-1}+\bar{\Lambda}G\bar{\Lambda}^T, \quad \Omega_{{N}\times N}=\hat{G}^{-1}+\Lambda^TH\Lambda, \\
&\tilde{\Omega}^a=\left(\begin{array}{cc}
0&\tilde{E}\\
\tilde{E}^T&\tilde{\Omega}\\
\end{array}\right), \quad {\Omega}^a=\left(\begin{array}{cc}
0&{E}\\
{E}^T&{\Omega}\\
\end{array}\right).
\end{aligned}
\end{equation}
with $\Lambda$ is the $\tilde{N}\times N$ Cauchy type matrix with $\Lambda_{lj}=\frac{1}{\lambda_l+\bar{k}_j}$. Here
\begin{equation}\label{de6}
\begin{aligned}
&\tilde{E}=(1,1,...,1)_{\tilde{N}}, \quad E=(1,1,...,1)_N, \\
&G={\rm diag}(g_1,g_2,\cdots,g_N), \quad H={\rm diag}(h_1,h_2,\cdots,h_{\tilde{N}}),
\end{aligned}
\end{equation}
and the diagonal matrices $\hat{G}, \hat{H}$ can be defined similarly from $\{\hat{g}_j\}$ and $\{\hat{h}_l\}$.
It is important to express the solution of the cnNLS equation in the form (\ref{e2}),(\ref{e3}) and (\ref{de6}).
Because it makes us possible and easier to give the general nonlocal reduction of the cnNLS equation, which will be discussed in the next section.

For $N=\tilde{N}=1$,
\begin{equation}\label{e4}
\begin{aligned}
q(x,t)&=\frac{2i}{\hat{h}_1^{-1}%\frac{1}{\bar{d}_1}\mathrm e^{2i\theta(-\bar\lambda_1;x,t)}
+\frac{g_1}{(\bar\lambda_1+k_1)^2}},\\%\mathrm e^{2i\theta(k_1;x,t)}},\\
\hat{q}(x,t)&=\frac{-2i}{\hat{g}_1^{-1}%\frac{1}{\bar{c}_1}\mathrm e^{2i\theta(-\bar{k}_1;x,t)}
+\frac{h_1}{(\lambda_1+\bar{k}_1)^2}},%\mathrm e^{2i\theta(\lambda_1;x,t)}}.
\end{aligned}
\end{equation}
where $\hat{g}_j,\hat{h}_j$ are defined in (\ref{e1}). If $k_1$ and $\lambda_1$ are chosen to be imaginary numbers, then the wave trains of $q(x,t)$ and $\hat{q}(x,t)$ are on different lines (see Figure \ref{fi1}).
\begin{figure}[h]\label{fi1}
  \centering
  \includegraphics[width=5cm,height=5cm]{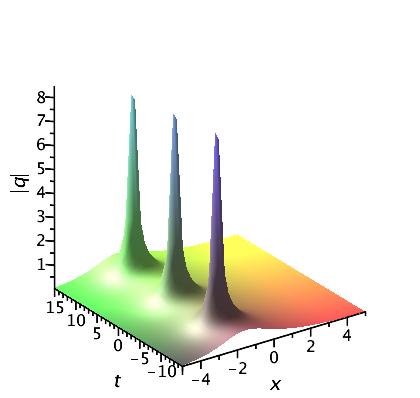} \quad
   \includegraphics[width=4.5cm,height=4.5cm]{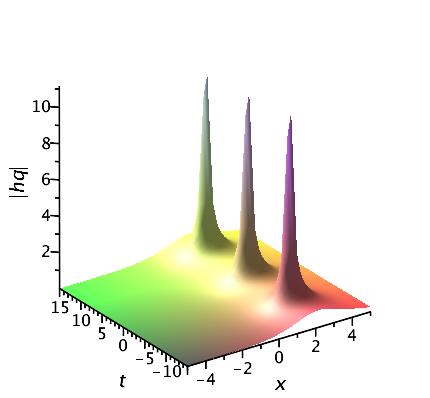}
  \caption{The amplitude of $q(x,t)$ (left) and $\hat{q}(x,t)$ (right) in (\ref{e4}) with $\lambda_1=0.5i,k_1=-0.3i, c_1={\rm e}^{1+i\pi/3}, d_1={\rm e}^{1.5+i\pi/4}$.
  }\label{fi1}
\end{figure}

For $N=1, \tilde{N}=2$, we have the solution (\ref{e2}) with
\begin{equation}\label{d7}
\begin{aligned}
\det\tilde\Omega^a=-\left[\hat{h}_1^{-1}+\hat{h}_2^{-1}%\frac{{\rm e}^{2i\theta(-\bar\lambda_1)}}{\bar{d}_1}
%+\frac{{\rm e}^{2i\theta(-\bar\lambda_2)}}{\bar{d}_2}
+\frac{g_1(\bar\lambda_2-\bar\lambda_1)^2}%{\rm e}^{2i\theta(k_1)}}
{(\bar\lambda_1+k_1)^2(\bar\lambda_2+k_1)^2}
\right],\\
\det\tilde{\Omega}=\hat{h}_1^{-1}\hat{h}_2^{-1}+\frac{g_1\hat{h}_1^{-1}}
{(\bar\lambda_2+k_1)^2}
+\frac{g_1\hat{h}_2^{-1}}{(\bar\lambda_1+k_1)^2}.
\end{aligned}
\end{equation}
and $\det\Omega^a=-1$,
\begin{equation}\label{d8}
 \det\Omega=\hat{g}_1^{-1}%\frac{{\rm e}^{2i\theta(-\bar{k}_1)}}{\bar{c}_1}
+\frac{h_1}%{\rm e}^{2i\theta(\lambda_1)}}
{(\lambda_1+\bar{k}_1)^2}
+\frac{h_2}%{\rm e}^{2i\theta(\lambda_2)}}
{(\lambda_2+\bar{k}_1)^2}.
 \end{equation}
Figure \ref{fi2} shows the amplitude of $q(x,t)$  and $\hat{q}(x,t)$ in (\ref{d7}),(\ref{d8}) and (\ref{e2}) with
$\lambda_1, \lambda_2$ and $k_1$ are imaginary numbers.
\begin{figure}[h]\label{fi2}
  \centering
  \includegraphics[width=5cm,height=5cm]{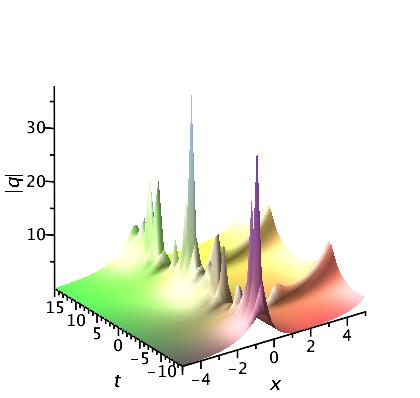} \quad
   \includegraphics[width=4.5cm,height=4.5cm]{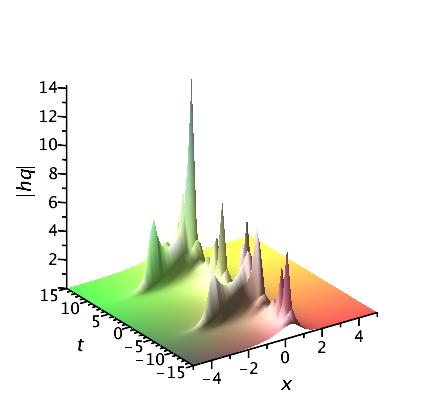}
  \caption{The amplitude of $q(x,t)$ (left) and $\hat{q}(x,t)$ (right) in (\ref{d7}),(\ref{d8}) and (\ref{e2}) with $\lambda_1=0.5i,\lambda_2=0.6i,k_1=-2i, c_1={\rm e}^{1+i\pi/3}, d_1={\rm e}^{1.5+i\pi/4},d_2={\rm e}^{1.2+i\pi/6}$.
  }\label{fi2}
\end{figure}

\begin{figure}[h]
  \centering
  \includegraphics[width=5cm,height=5cm]{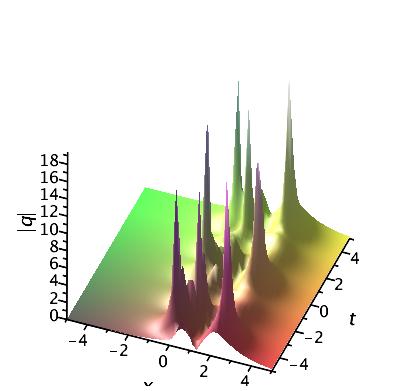} \quad
   \includegraphics[width=4.5cm,height=4.5cm]{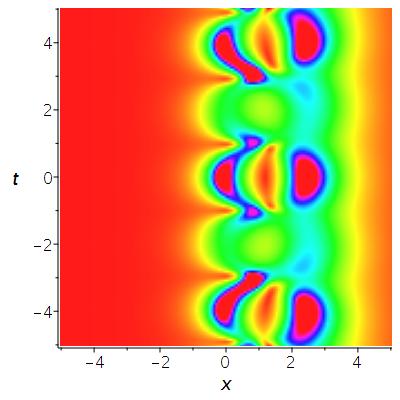}
  \caption{The amplitude of $q(x,t)$ and its density in (\ref{e12}) and (\ref{e2}) with $\lambda_1=0.8i,\lambda_2=1.2i,k_1=-i,k_2=-0.5i, c_1=c_2=1, d_1=d_2=1$.
  }\label{fi3}
\end{figure}

\begin{figure}[h]
  \centering
  \includegraphics[width=5cm,height=5cm]{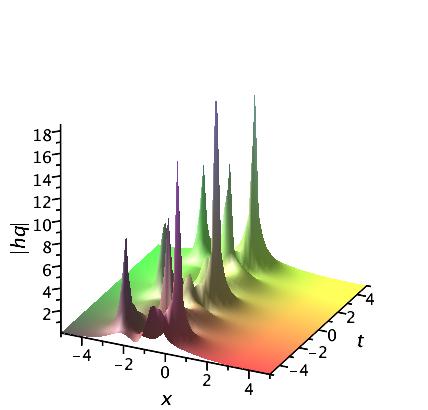} \quad
   \includegraphics[width=4.5cm,height=4.5cm]{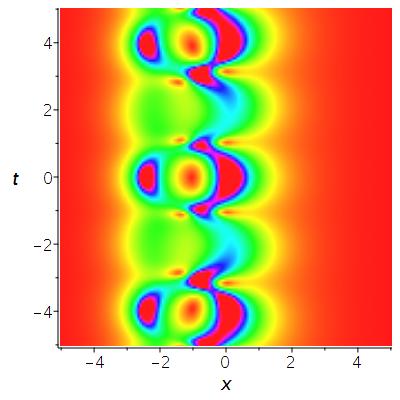}
  \caption{The amplitude of $\hat{q}(x,t)$ and its density in (\ref{e13}) and (\ref{e2}) with $\lambda_1=0.8i,\lambda_2=1.2i,k_1=-i,k_2=-0.5i, c_1=c_2=1, d_1=d_2=1$.
  }\label{fi4}
\end{figure}

Similarly, for $N=\tilde{N}=2$, we obtain the solution (\ref{e2}) with
\begin{equation}\label{e12}
\begin{aligned}
\det\tilde\Omega^a=&-\left[\hat{h}_1^{-1}+\hat{h}_2^{-1}+\frac{g_1(\bar\lambda_2-\bar\lambda_1)^2}
{(\bar\lambda_1+k_1)^2(\bar\lambda_2+k_1)^2}+\frac{g_2(\bar\lambda_2-\bar\lambda_1)^2}%{\rm e}^{2i\theta(k_1)}}
{(\bar\lambda_1+k_2)^2(\bar\lambda_2+k_2)^2}\right],\\
\det\tilde\Omega=&\hat{h}_1^{-1}\hat{h}_2^{-1}+g_1g_2\frac{(\bar\lambda_2-\bar\lambda_1)^2(k_2-k_1)^2}{\prod_{j,l=1}^2(\bar\lambda_j+k_l)^2}\\
&+\frac{g_1\hat{h}_1^{-1}}{(\bar\lambda_2+k_1)^2}+\frac{g_2\hat{h}_1^{-1}}{(\bar\lambda_2+k_2)^2}
+\frac{g_1\hat{h}_2^{-1}}{(\bar\lambda_1+k_1)^2}+\frac{g_2\hat{h}_2^{-1}}{(\bar\lambda_1+k_2)^2},
\end{aligned}
\end{equation}
and
\begin{equation}\label{e13}
\begin{aligned}
\det\Omega^a=&-\left[\hat{g}_1^{-1}+\hat{g}_2^{-1}+\frac{h_1(\bar{k}_2-\bar{k}_1)^2}
{(\lambda_1+\bar{k}_1)^2(\lambda_1+\bar{k}_2)^2}+\frac{h_2(\bar{k}_2-\bar{k}_1)^2}
{(\lambda_2+\bar{k}_1)^2(\lambda_2+\bar{k}_2)^2}\right],\\
\det\Omega=&\hat{g}_1^{-1}\hat{g}_2^{-1}+h_1h_2\frac{(\lambda_2-\lambda_1)^2(\bar{k}_2-\bar{k}_1)^2}{\prod_{j,l=1}^2(\lambda_j+\bar{k}_l)^2}\\
&+\frac{h_1\hat{g}_1^{-1}}{(\lambda_1+\bar{k}_2)^2}+\frac{h_1\hat{g}_2^{-1}}{(\lambda_1+\bar{k}_1)^2}
+\frac{h_2\hat{g}_1^{-1}}{(\lambda_2+\bar{k}_2)^2}
+\frac{h_2\hat{g}_2^{-1}}{(\lambda_2+\bar{k}_1)^2}.
\end{aligned}
\end{equation}
Figure \ref{fi3} and Figure \ref{fi4} show the amplitude of $q(x,t)$ and $\hat{q}(x,t)$ in (\ref{e12}),(\ref{e13}) and (\ref{e2}) with $\lambda_1, \lambda_2$ and $k_1,k_2$ are imaginary numbers.

From the above figures, we find that the two wave trains of $q(x,t)$ and $\hat{q}(x,t)$ locate on different lines which are determined by the Dbar data $\{\lambda_l, d_l\}$ and $\{k_j,c_j\}$. So we can carefully choose certain Dbar data to ensure $q(x,t)=\hat{q}(x,t)$.
%where $\tilde{M}=I+A\hat{B}$ is a $\tilde{N}\times \tilde{N}$ matrix and  ${M}=I+B\hat{A}$ is a $N\times N$ matrix
%\begin{equation}\label{f6}
%\begin{aligned}
% \quad \tilde{M}^a=\left(\begin{array}{cc}
%0&\hat{h}\\
%\tilde{E}^T&\tilde{M}\\
%\end{array}\right), \quad {M}^a=\left(\begin{array}{cc}
%0&\hat{g}\\
%{E}^T&{M}\\
%\end{array}\right).\\
%\end{aligned}
%\end{equation}
%Here
%\begin{equation}\label{d6}
%\begin{aligned}
%&\hat{g}=(\hat{g}_1,\hat{g}_2,...,\hat{g}_N),\quad \hat{h}=(\hat{h}_1,\hat{h}_2,...,\hat{h}_{\tilde{N}}), \\
%&E=(1,1,...,1)_N, \quad \tilde{E}=(1,1,...,1)_{\tilde{N}},\\
%\end{aligned}
%\end{equation}
%and the elements of the $\tilde{N}\times {N}$ matrices $A,\hat{A}$ and ${N}\times \tilde{N}$ matrices $B,\hat{B}$
%take the following form
%\begin{equation}\label{d6b}
%\begin{aligned}
%%&H=(H_{jm}),\quad G=(G_{jm}), \quad\tilde{G}(x)=\overline{G(-x)}, \quad \tilde{H}(x)=\overline{H(-x)}\\%=\mathrm e^{2(\tilde{\eta}_j+i\tilde{\phi}_j)},\\
%&A_{lj}=\frac{g_j}{\bar{\lambda}_l+{k}_j}, \quad \hat{A}_{lj}=\frac{\hat{g}_j}{{\lambda}_l+\bar{k}_j},\\
%%\quad {h}_j={d}_j\mathrm e^{2i\theta({\lambda}_j;x,t)},\quad %\bar{c}_j=\mathrm e^{-2(\alpha_j+i\beta_j)},\quad \bar{d}_j=\mathrm e^{2(\tilde{\alpha}_j+i\tilde{\beta}_j)},\\
% &B_{jl}=\frac{h_l}{\bar{k}_j+{\lambda}_l}, \quad \hat{B}_{jl}=\frac{\hat{h}_l}{{k}_j+\bar{\lambda}_l}.
%%&&g_j=\mathrm e^{-2(\tilde{\eta}_j+i\tilde{\phi}_j)},\quad h_j=\mathrm e^{2(\eta_j+i\phi_j)},\\
%\end{aligned}
%\end{equation}

\section{Reduction to the nonlocal focusing NLS equation}\vspace{0.3cm}
In general, $q(x,t)$ and $\hat{q}(x,t)$ given in (\ref{e2}) are not equal to each other.
From (\ref{e1}), we find that $q(x,t)=\hat{q}(x,t)$ implies the following conditions $k_j=-\bar{k}_j$, $\lambda_l=-\bar\lambda_l$ and $N=\tilde{N}$, as well as $|c_j|, |d_l|$ dependenting on $\{k_j,\lambda_l\}$.
To find the constraint conditions on $|c_j|$ and $|d_l|$ for $q(x,t)=\hat{q}(x,t)$, we take $k_j=ib_{j},~\lambda_l=i\eta_{l},$ with $b_j<0<\eta_l, j,l=1,2,\cdots,N $. It is remarked that
\begin{equation}\label{d6}
 \begin{aligned}
g_j=c_j{\rm e}^{2i\theta(ib_j)}, \quad \hat{g}_j^{-1}=\frac{1}{\bar{c}_j}{\rm e}^{2i\theta(ib_j)},\\
h_j=d_j{\rm e}^{2i\theta(i\eta_j)}, \quad \hat{h}_j^{-1}=\frac{1}{\bar{d}_j}{\rm e}^{2i\theta(i\eta_j)}.
\end{aligned}
\end{equation}

Since the matrices $\Omega$ and $\tilde\Omega$ in (\ref{e3}) are symmetry matrix and the matrix $K=i\Lambda=(\frac{1}{\eta_j-b_l})$ is a Cauchy matrix. Using the Cauchy-Binet formula, we give the determinants in (\ref{e2}) as the following form \cite{jpa46-035204,mpag17-49}
\begin{equation}\label{e5}
\begin{array}{c}
 \det\tilde\Omega^a=\sum\limits_{\sigma=1}^N(-1)^\sigma\sum\limits_{\sigma12}\prod\limits_{l,s,n_\sigma}\frac{\hat{h}_{n_\sigma}^{-1}g_s}{(\eta_l-b_s)^2}
\prod\limits_{\tiny\begin{array}{c}
l<l'\\[-3pt]
s<s'
\end{array}}(\eta_l-\eta_{l'})^2(b_{s'}-b_s)^2,\\
\det\tilde\Omega=\prod\limits_{j=1}^N\hat{h}_j^{-1}+\sum\limits_{\sigma=1}^N(-1)^\sigma\sum\limits_{\sigma11}\prod\limits_{l,m,n_\sigma}\frac{\hat{h}_{n_\sigma}^{-1}g_m}{(\eta_l-b_m)^2}
\prod\limits_{\tiny\begin{array}{c}
l<l'\\[-3pt]
m<m'
\end{array}}(\eta_l-\eta_{l'})^2(b_{m'}-b_m)^2,\\
%l,l'\in\{j_1,j_2,\cdots,j_\sigma\}, \quad  s,s'\in\{r_2,\cdots,r_\sigma\}, \quad m,m'\in\{r_1,r_2,\cdots,r_\sigma\} ,\\
%n_\sigma\in\{1,2,\cdots,N\}\backslash\{j_1,\cdots,j_\sigma\},
\end{array}
\end{equation}
and
\begin{equation}\label{e5a}
\begin{array}{c}
 \det\Omega^a=\sum\limits_{\sigma=1}^N(-1)^\sigma\sum\limits_{\sigma12}\prod\limits_{l,s,n_\sigma}\frac{h_s\hat{g}_{n_\sigma}^{-1}}{(\eta_s-b_l)^2}
\prod\limits_{\tiny\begin{array}{c}
l<l'\\[-3pt]
s<s'
\end{array}}(\eta_s-\eta_{s'})^2(b_{l'}-b_l)^2,\\
\det\Omega=\prod\limits_{j=1}^N\hat{g}_j^{-1}+\sum\limits_{\sigma=1}^N(-1)^\sigma\sum\limits_{\sigma11}\prod\limits_{l,m,n_\sigma}\frac{h_m\hat{g}_{n_\sigma}^{-1}}{(\eta_m-b_l)^2}
\prod\limits_{\tiny\begin{array}{c}
l<l'\\[-3pt]
m<m'
\end{array}}(\eta_m-\eta_{m'})^2(b_{l'}-b_l)^2,\\
l,l'\in\{j_1,j_2,\cdots,j_\sigma\}, \quad  s,s'\in\{r_2,\cdots,r_\sigma\}, \quad m,m'\in\{r_1,r_2,\cdots,r_\sigma\} ,\\
n_\sigma\in\{1,2,\cdots,N\}\backslash\{j_1,\cdots,j_\sigma\},
\end{array}
\end{equation}
where $\sum\limits_{\sigma12}$ denotes the summation for indices $1\leq j_1\leq j_2\leq\cdots\leq j_\sigma\leq N,
1\leq r_2\leq\cdots\leq r_\sigma\leq N$, and summation $\sum\limits_{\sigma11}$ for $1\leq j_1\leq j_2\leq\cdots\leq j_\sigma\leq N, 1\leq r_1\leq r_2\leq\cdots\leq r_\sigma\leq N$.

If the determinants in (\ref{e2}) admit
\begin{equation}\label{d17}
\det\tilde\Omega^a=(-1)^{N-1}\det\Omega^a, \quad \det\tilde\Omega=(-1)^N\det\Omega,
\end{equation}
which means that $\det\tilde\Omega^a\det\Omega+\det\Omega^a\det\tilde\Omega=0$, then $q(x,t)=\hat{q}(x,t)$, and the cnNLS equation (\ref{a1}) reduces to the nNLS equation (\ref{a1a}). In addtion, from (\ref{e5}), (\ref{e5a}) and  (\ref{d17}), the constraint conditions about $|c_j|, |d_j|$ on  $\eta_j$ and $b_j$ can be derived.
%then (\ref{d7}) can be rewritten as
%\begin{equation}\label{e4}
%q(x,t)=-2i\frac{\det \tilde{\Omega}^a}{\det{ \tilde\Omega}}, \quad \hat{q}(x,t)=2i\frac{\det {{\Omega}}^a}{\det{ {\Omega}}},
%\end{equation}
%where the symmetry matrix $\tilde\Omega$ and $\Omega$ are given by
%\begin{equation}\label{e5}
%\begin{aligned}
%&\tilde\Omega=\hat{H}^{-1}-KGK^T, \quad \Omega=\hat{G}^{-1}-K^THK, \\
%&\tilde{\Omega}^a=\left(\begin{array}{cc}
%0&{E}\\
%{E}^T&\tilde{\Omega}\\
%\end{array}\right), \quad {\Omega}^a=\left(\begin{array}{cc}
%0&{E}\\
%{E}^T&{\Omega}\\
%\end{array}\right).
%\end{aligned}
%\end{equation}
%with $K$ is the Cauchy matrix $K_{jl}=\frac{1}{\eta_j-b_l}$. Here $ G={\rm diag}(g_1,g_2,\cdots,d_N), \quad H={\rm diag}(h_1,h_2,\cdots,h_N)$, and the diagonal matrices $\hat{G}, \hat{H}$ have the same definition.

\begin{figure}[htbp]
 %\centering
  \begin{minipage}[t]{0.45\textwidth}
  \centering
  \includegraphics[width=4.5cm,height=4.5cm]{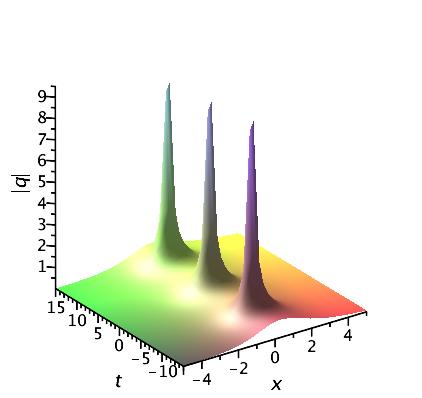}
  \caption{The amplitude of $q(x,t)$ in (\ref{d8}) with $\eta_1=0.5$, $b_1=-0.3, d_1=0.8
  {\rm e}^{i\pi/4}, c_1=0.8{\rm e}^{i \pi/3}$.}\label{fi5}
 \end{minipage}\qquad
 \begin{minipage}[t]{0.45\textwidth}
 \centering
  \includegraphics[width=5cm,height=5cm]{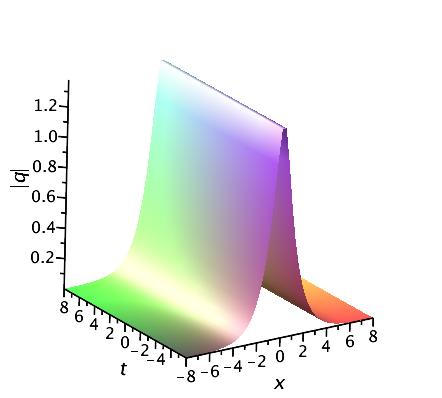}
  \caption{The amplitude of $q(x,t)$ in (\ref{d8}) with $\eta_1=\frac{1}{2}$, $b_1=-\frac{1}{2}, d_1=
  {\rm e}^{i2\pi/3}, c_1={\rm e}^{i\pi/6}$.}\label{fi6}
  \end{minipage}
  %\begin{minipage}[t]{0.45\textwidth}
%  \centering
%  \includegraphics[width=5cm,height=5cm]{n1rr}
%  \caption{The amplitude of $q(x,t)$ in (\ref{d8}) with $\eta_1=\frac{1}{2}$, $b_1=-\frac{1}{2}, d_1=
%  {\rm e}^{i 2\pi/3}, c_1={\rm e}^{i \pi/3}$.}
% \end{minipage}\qquad\qquad
%   \begin{minipage}[t]{0.45\textwidth}
%  \centering
%  \includegraphics[width=5cm,height=5cm]{n1r2}
%  \caption{The amplitude of $q(x,t)$ in (\ref{d8}) with $\eta_1=\frac{1}{2}$, $b_1=-\frac{1}{2}, d_1=
%  {\rm e}^{i \pi/3}, c_1={\rm e}^{i 4\pi/3}$.}
% \end{minipage}
 % \caption{Solution (\ref{d8}) with $\rho=e^{i\frac{\pi}{4}}, q_0=1, \alpha_1=\frac{\pi}{3}, \epsilon_j=0$.}
\end{figure}

For $N=\tilde{N}=1$, We have, from (\ref{e5}), (\ref{e5a}), $\det\tilde\Omega^a=\det\Omega^a=-1$ and
$$\det\tilde\Omega=\hat{h}_1^{-1}-\frac{1}{(\eta_1-b_1)^2}g_1, \quad \det\Omega=\hat{g}_1^{-1}-\frac{1}{(\eta_1-b_1)^2}h_1,$$
which can also be obtained from (\ref{e4}). Then $\det\tilde\Omega=-\det\Omega$ implies that
\begin{equation}\label{e6}
 |c_1|=|d_1|=\eta_1-b_1,
\end{equation}
in terms of (\ref{d6}).
In this case, we take $c_1={\rm e}^{2(r_1+i\alpha_1)}, d_1={\rm e}^{2(r_1+i\beta_1)}$, then the solution of nNLS equation takes the form of
\begin{equation}\label{e7}
 q(x,t)=i\frac{{\rm e}^{2X}{\rm e}^{-2i\tilde{T}}-{\rm e}^{2\tilde{X}}{\rm e}^{-2i{T}}}{\cosh[2(\tilde{X}-X)]-\cos[2(\tilde{T}-T)]},
\end{equation}
where $\eta_1-b_1={\rm e}^{2r_1}$ and
$$\tilde{X}=\eta_1x+r_1, \quad X=b_1x+r_1, \quad \tilde{T}=2\eta_1^2t+\beta_1, \quad T=2b_1^2t+\alpha_1.$$
This solution is singular at the points (see Figure \ref{fi5})
\begin{equation}\label{e8}
 x=0, \quad t=\frac{n\pi+\alpha_1-\beta_1}{2(\eta_1^2-b_1^2)}, \quad n\in\mathbb{Z}.
\end{equation}

%In particular, we take $k_j=ib_{j},~\lambda_j=i\eta_{j}$ and consider the case $N=1$, then stationary solution is obtained in the following form
%\begin{equation}\label{d8}
%q(x,t)=\frac{(b_1-\eta_1)\big({\rm e}^{2X}{\rm e}^{-2i\tilde{T}}+{\rm e}^{2\tilde{X}}{\rm e}^{-2i{T}}\big)}
%{\cosh2(X-\tilde{X})+\cos2(T-\tilde{T})},
%\end{equation}
%where we take $\eta_1>0>b_1$ and
%\begin{equation}\label{d8a}
%\begin{aligned}
%X=\eta_1x+x_0, \quad T=2\eta_1^2t+t_0, \\
%\tilde{X}=b_1x+\tilde{x}_0, \quad \tilde{T}=2b_1^2t+\tilde{t}_0,\\
%x_0=-\frac{1}{2}\ln\frac{|d_1|}{\eta_1-b_1}, \quad t_0=\frac{1}{2}\arg d_1+\frac{\pi}{4},\\
%\tilde{x}_0=\frac{1}{2}\ln\frac{|c_1|}{\eta_1-b_1}, \quad \tilde{t}_0=\frac{1}{2}\arg c_1-\frac{\pi}{4}.\\
%\end{aligned}
%\end{equation}
%We note that $|q|$ rapidly decay to zero as $|x|\to\infty$ implies that $\eta_1b_1<0$.
%%It is noted that the solution (\ref{d8}) has no singularities.
%%In the case of $|\eta_1|\neq |b_1|$,
%When $\eta_1\neq-b_1$, the solution (\ref{d8}) %is periodic on time.
%has some singularities at points (see Fig. 1)
%\begin{equation}\label{d8b}
%\begin{aligned}
% x=&\frac{1}{2(\eta_1-b_1)}\ln\frac{|c_1d_1|}{(\eta_1-b_1)^2}, \\
% t=&\frac{2n\pi+\arg c_1-\arg d_1}{4(\eta_1^2-b_1^2)}, \quad n\in\mathbb{Z}.
%\end{aligned}
%\end{equation}
In particular, if $b_1=-\eta_1$, then $\tilde{T}-T=\beta_1-\alpha_1$. Furthermore, if $\cos[2(\tilde{T}-T)]=1$, ($\beta_1=\alpha_1$ or $\beta_1=\alpha_1+\pi$), solution of the nNLS equation (\ref{a1a}) reduces to
\begin{equation}\label{d8d}
q(x,t)=-2i\eta_1{\rm e}^{-2i\tilde{T}}{\rm csch}(2\eta_1x),
\end{equation}
which is singular at the line $x=0$.

In the case $b_1=-\eta_1$ and $|\cos[2(\tilde{T}-T)]|<1$, (\ref{e7}) gives the stationary soliton solution of the nNLS equation (\ref{a1a}) (see Fighure \ref{fi6}). While in the case $b_1=-\eta_1$ and $\cos[2(\tilde{T}-T)]=-1$ or $\beta_1=\alpha_1\pm\pi/2$, (\ref{e7}) yields
\begin{equation}\label{d8c}
q(x,t)=2i\eta_1{\rm e}^{-2i\tilde{T}}{\rm sech}(2\eta_1x),
\end{equation}
which means that the two distributions $R(k,x,t)$ and $\hat{R}(k;x,t)$ are equal. Hence, the nNLS equation (\ref{a1a}) reduces to the NLS equation.

%(\ref{d8}) reduces to another solution (see Fig. 2)

%While in the case of $\eta_1=-b_1$ and $\arg c_1\neq\arg d_1$, that is $-1<\cos2(T-\tilde{T})\leq1$, the soliton (\ref{d8}) (the case of $-1<\cos2(T-\tilde{T})<1$ is shown in Fig. 3 and the case of $\cos2(T-\tilde{T})=1$ is shown in Fig. 4). It is noted that the solution (\ref{d8}) will reduce to
%$$q(x,t)=-2\eta_1{\rm e}^{2iT}{\rm sech}(2X),$$
%as $\eta_1=-b_1$, $\arg d_1=\arg c_1\pm\pi$ and $|d_1|=|c_1|$, which means that $\overline{q(-x,t)}=\overline{q(x,t)}$.

%\begin{figure}[htbp]
% %\centering
%  % Requires \usepackage{graphicx}
%  \begin{minipage}[t]{0.45\textwidth}
%  \centering
%  \includegraphics[width=4.5cm,height=4.5cm]{n2}
%  \caption{Solution (\ref{d13}) with $\eta_1=0.5, \eta_2=1, b_2=-0.5, b_1=-1$.}
% \end{minipage}\qquad\qquad
%\begin{minipage}[t]{0.45\textwidth}
% \centering
%  \includegraphics[width=5cm,height=5cm]{n2s}
%  \caption{Solution (\ref{d13}) with $\eta_1=1, \eta_2=2, b_2=-0.5, b_1=-1$.}
%  \end{minipage}
% % \caption{Solution (\ref{d8}) with $\rho=e^{i\frac{\pi}{4}}, q_0=1, \alpha_1=\frac{\pi}{3}, \epsilon_j=0$.}
%\end{figure}

%From equation (\ref{d7}), if $k_{1,R}=\lambda_{1,R}=0$,
%\begin{equation}\label{d9}
%\begin{array}{rcl}
%&&\tilde{\phi}_1-{\phi}_1=2(-k^2_{1,I}+\lambda^2_{1,I})t+\tilde{\beta}_1-\beta_1,\\
%\end{array}
%\end{equation}
%when $t$ from $0$ to $\infty$, $m_1$ will alternate positive and negative variations.
%
%Hence, when $\sin2(\tilde{\phi}_1-{\phi}_1)=0$, $m_1>0$, we obtain the bright soliton solution
%\begin{equation}\label{d10}
%\begin{aligned}
%&q(x,t)=i\mathrm e^{2i\phi_1}\mathrm e^{-({\eta}_1+\tilde{\eta}_1+\varphi_1)} \sech{(\tilde{\eta}_1-{\eta}_1+\varphi_1)},\\
%&m_1=\mathrm e^{2\varphi_1},\\
%\end{aligned}
%\end{equation}
%when $\sin2(\tilde{\phi}_1-{\phi}_1)=0$, $m_1<0$, we obtain the dark soliton solution
%\begin{equation}\label{d11}
%\begin{aligned}
%&q(x,t)=-i\mathrm e^{2i\phi_1}\mathrm e^{-({\eta}_1+\tilde{\eta}_1+\varphi_1)}\mathrm {csch} {(\tilde{\eta}_1-{\eta}_1+\varphi_1)},\\
%&m_1=-\mathrm e^{2\varphi_1}.\\
%\end{aligned}
%\end{equation}

For $N=\tilde{N}=2$, using (\ref{e5}), (\ref{e5a}) or equivalent (\ref{e12}) and (\ref{e13}), we find, from $\det\tilde\Omega^a=-\det\Omega^a$, that
\begin{equation}\label{e9}
 |c_j|=\frac{(\eta_1-b_j)(\eta_2-b_j)}{|\eta_2-\eta_1|}, \quad
 |d_j|=\frac{(\eta_j-b_1)(\eta_j-b_2)}{|b_2-b_1|}, \quad j=1,2,
\end{equation}
and $\det\tilde\Omega=\det\Omega$ implies that $|c_1c_2|=|d_1d_2|$
%then $q(x,t)=\hat{q}(x,t)$. We note that the condition (\ref{e9})
and
\begin{equation}\label{d9}
\frac{|c_1|^2}{|d_1|^2}=\frac{(\eta_2-b_1)^2}{(\eta_1-b_2)^2}=\frac{|d_2|^2}{|c_2|^2}, \quad
\frac{|c_2|^2}{|d_1|^2}=\frac{(\eta_2-b_2)^2}{(\eta_1-b_1)^2}=\frac{|d_2|^2}{|c_1|^2}.
\end{equation}
In addition, $|c_1c_2|=|d_1d_2|$ and (\ref{e9}) give $|\eta_2-\eta_1|=|b_2-b_1|$.
 %in (\ref{d6}), we let ${\lambda}_j=i\eta_j, {k}_j=i{b}_j, (j=1,2)$. We choose $b_j<0$ and $\eta_j>0, (j=1,2)$ % in view of the boundary condition at $|x|=\infty$.

\begin{figure}[h]
  \centering
  \includegraphics[width=5cm,height=5cm]{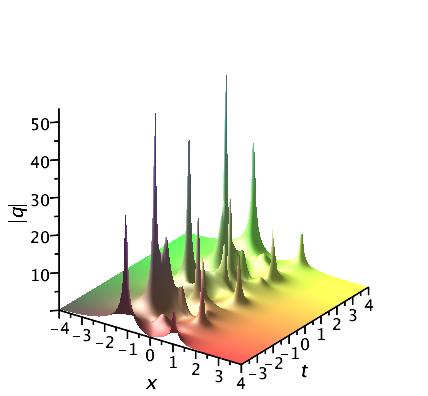} \quad
   \includegraphics[width=4.5cm,height=4.5cm]{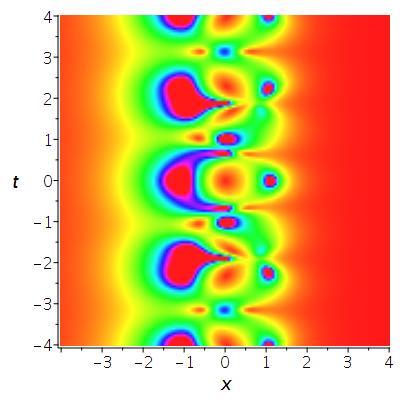}
  \caption{The amplitude and its density of $q(x,t)$ in (\ref{d20}) with $\eta_1=0.5,\eta_2=1, b_1=-0.8,b_2=-0.3, \alpha_j=0, \beta_j=0, (j=1,2)$.
  }\label{fi9}
\end{figure}

\begin{figure}[h]
  \centering
  \includegraphics[width=5cm,height=5cm]{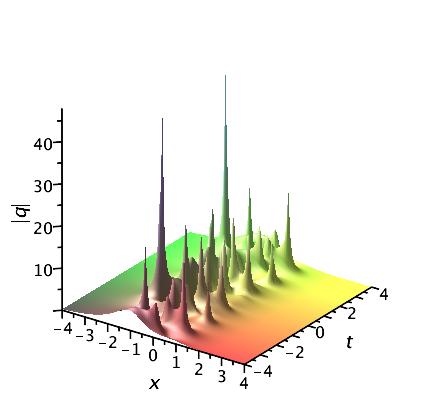} \quad
   \includegraphics[width=4.5cm,height=4.5cm]{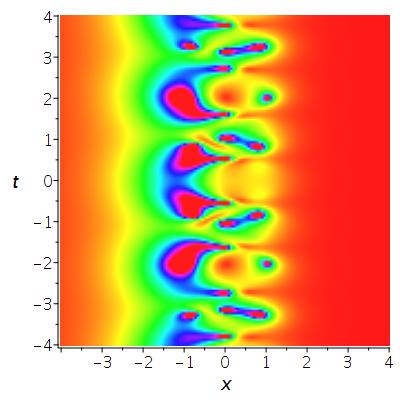}
  \caption{The amplitude and its density of $q(x,t)$ in (\ref{d20}) with $\eta_1=0.5,\eta_2=1, b_1=-0.8,b2=-1.3, \alpha_j=0, \beta_j=0, (j=1,2)$.
  }\label{fi10}
\end{figure}

For convenience, we let $c_j=|c_j|{\rm e}^{2i\alpha_j}, d_j=|d_j|{\rm e}^{2i\beta_j}$ and
\begin{equation}\label{d12}
\begin{aligned}
g_j=|c_j|\mathrm e^{2{\theta}_j},\quad \hat{g}_j^{-1}=\frac{1}{|c_j|}\mathrm e^{2{\theta}_j},\\
h_j=|{d}_j|\mathrm e^{2\tilde\theta_j}, \quad \hat{h}_j^{-1}=\frac{1}{|d_j|}\mathrm e^{2{\tilde\theta}_j},
\end{aligned}
\end{equation}
where
\begin{equation}\label{d13}
\begin{aligned}
\theta_j=-X_j+iT_j, \quad X_j=b_jx, \quad T_j=2b_j^2t+{\alpha_j},\\
\tilde\theta_j=-\tilde{X}_j+i\tilde{T}_j, \quad \tilde{X}_j=\eta_jx, \quad \tilde{T}_j=2\eta_j^2t+{\beta_j}.\\
\end{aligned}
\end{equation}
%and
%\begin{equation}\label{d9}
%\begin{aligned}
%\tilde{X}_j=b_jx, \quad \tilde{T}_j=2b_j^2t+\tilde{t}_{j0}, \quad \frac{\bar{c}_j}{\tilde{m}_j}=-{\rm e}^{-2i\tilde{t}_{j0}},\\
%X_j=\eta_jx,\quad T_j=2\eta_j^2t+t_{j0}, \quad \frac{d_j}{m_j}={\rm e}^{2it_{j0}}.
%\end{aligned}
%\end{equation}
Then we find that solution of nNLS equation (\ref{a1a}) is
\begin{equation}\label{e10}
 q(x,t)=-2i\frac{ M^a}{ M},
\end{equation}
with
\begin{equation}\label{d11}
 \begin{aligned}
 M^a=|c_2|{\rm e}^{2\theta_1}+|c_1|{\rm e}^{2\theta_2}
-|d_2|{\rm e}^{2\tilde\theta_1}-|d_1|{\rm e}^{2\tilde\theta_2},
 \end{aligned}
\end{equation}
\begin{equation}\label{d10}
\begin{aligned}
 M=&{\rm e}^{2(\tilde\theta_1+\tilde\theta_2)}+{\rm e}^{2(\theta_1+\theta_2)}\\
&-p_1\big({\rm e}^{2(\tilde\theta_1+\theta_2)}+{\rm e}^{2(\tilde\theta_2+\theta_1)}\big)
-p_2\big({\rm e}^{2(\tilde\theta_1+\theta_1)}+{\rm e}^{2(\tilde\theta_2+\theta_2)}\big).
\end{aligned}
\end{equation}
where $p_j$ are constants and satisfy the following relations
\begin{equation}\label{d15}
\begin{aligned}
p_1=&\frac{|d_1{c}_1|}{(\eta_1-b_1)^2}=\frac{|d_2c_2|}{(\eta_2-b_2)^2},\\
p_2=&\frac{|d_1c_2|}{(\eta_1-b_2)^2}=\frac{|d_2c_1|}{(\eta_2-b_1)^2}.
\end{aligned}
\end{equation}
We note that
\begin{equation}\label{d16}
 p_2-p_1=\frac{b_2-b_1}{\eta_2-\eta_1}={\rm sgn}(b_2-b_1),
\end{equation}
if $0<\eta_1<\eta_2$ in view of $|\eta_2-\eta_1|=|b_2-b_1|$. The solution (\ref{e10}) with $\eta_2-\eta_1=b_2-b_1$  is shown in Figure \ref{fi9}, and $\eta_2-\eta_1=b_1-b_2$ is shown in Figure \ref{fi10}.


\begin{figure}[h]
  \centering
  \includegraphics[width=5cm,height=5cm]{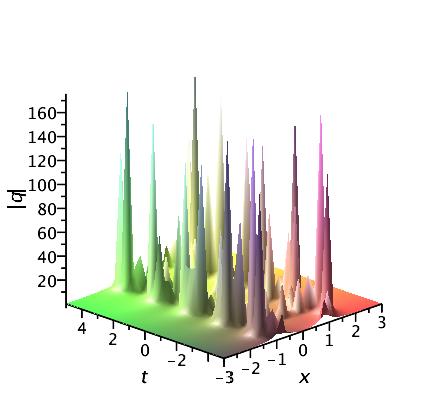} \quad
   \includegraphics[width=4.5cm,height=4.5cm]{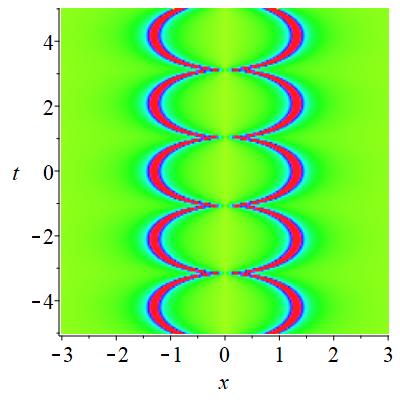}
  \caption{The amplitude and its density of $q(x,t)$ in (\ref{d19}) with $\eta_1=0.5,\eta_2=1, \alpha_j=0, \beta_j=0, (j=1,2)$.
  }\label{fi7}
\end{figure}

In particularly, if we take $\beta_{j}=\alpha_{j}$ and $b_j=-\eta_j, (j=1,2)$, then %$\tilde{X}_j=-X_j, \tilde{T}_j=T_j$ and
(\ref{e10}) reduces to
\begin{equation}\label{d18}
q(x,t)=\frac{-2i\big(m_1e^{-2i\tilde{T}_1}\sinh2\tilde{X}_2+m_2e^{-2i\tilde{T}_2}\sinh2\tilde{X}_1\big)}{\cosh2(\tilde{X}_2+\tilde{X}_1)-p\cosh2(\tilde{X}_2-\tilde{X}_1)-(p-1)\cos2(\tilde{T}_2-\tilde{T}_1)},
\end{equation}
where
$$m_1=\frac{2\eta_1(\eta_1+\eta_2)}{\eta_2-\eta_1}, \quad
m_2=\frac{2\eta_2(\eta_1+\eta_2)}{\eta_2-\eta_1}, \quad p=\frac{(\eta_2+\eta_1)^2}{(\eta_2-\eta_1)^2}>1.$$
and $\tilde{X}_j$ and $\tilde{T}_j$ are defined in (\ref{d12}). The solution (\ref{d19}) has some singularities, and is shown in Figure \ref{fi7}.

\begin{figure}[h]
  \centering
  \includegraphics[width=5cm,height=5cm]{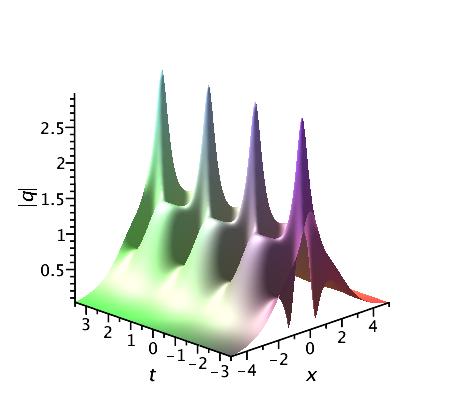} \quad
   \includegraphics[width=4.5cm,height=4.5cm]{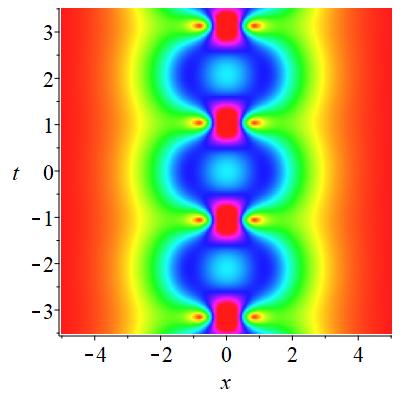}
  \caption{The amplitude and its density of $q(x,t)$ in (\ref{d20}) with $\eta_1=0.5,\eta_2=1, \alpha_j=0, \beta_j=\pi/2, (j=1,2)$.
  }\label{fi8}
\end{figure}

Now, if we take $b_1=-\eta_2, b_2=-\eta_1$ and
$ \beta_1=\alpha_2+\pi/2, \quad \beta_2=\alpha_1+\pi/2,$
then %$\tilde{X}_j=-X_j, \tilde{T}_j=T_j$ and
(\ref{e10}) reduces to a regular solution
\begin{equation}\label{d19}
q(x,t)=\frac{2i\big(m_1e^{-2i\tilde{T}_1}\cosh2\tilde{X}_2+m_2e^{-2i\tilde{T}_2}\cosh2\tilde{X}_1\big)}
{\cosh2(\tilde{X}_2+\tilde{X}_1)+(p_1+1)\cosh2(\tilde{X}_2-\tilde{X}_1)+p_1\cos2(\tilde{T}_2-\tilde{T}_1)},
\end{equation}
where
$$m_1=\frac{2\eta_1(\eta_1+\eta_2)}{\eta_2-\eta_1}, \quad
m_2=\frac{2\eta_2(\eta_1+\eta_2)}{\eta_2-\eta_1}, \quad p_1=\frac{4\eta_1\eta_2}{(\eta_2-\eta_1)^2}.$$
This solution is shown in Figure \ref{fi8}.

Furthermore, For $N=\tilde{N}=3$, one may find, from (\ref{e5}) and (\ref{e5a}), that
\begin{equation}\label{d20}
\begin{aligned}
\det\tilde\Omega^a=&-\sum\limits_{\tiny 1\leq s<s'\leq3}\hat{h}_s^{-1}\hat{h}_{s'}^{-1}+\sum\limits_{\tiny 1\leq l<l'\leq3}\sum\limits_{s=1}^3\frac{(\eta_l-\eta_{l'})^2}{(\eta_l-b_s)^2(\eta_{l'}-b_s)^2}g_s\hat{h}_{n}^{-1}\\
&-\sum\limits_{\tiny 1\leq s<s'\leq3}\frac{(b_{s'}-b_s)^2\prod\limits_{\tiny1\leq l<l'\leq3}(\eta_l-\eta_{l'})^2}{\prod\limits_{\tiny1\leq l\leq3}(\eta_l-b_s)^2(\eta_l-b_{s'})^2}g_sg_{s'}, \quad n=\{1,2,3\}\setminus\{l,l'\},
\end{aligned}
\end{equation}
\begin{equation}\label{d21}
\begin{aligned}
\det\Omega^a=&-\sum\limits_{\tiny 1\leq s<s'\leq3}\hat{g}_s^{-1}\hat{g}_{s'}^{-1}+\sum\limits_{\tiny 1\leq s<s'\leq3}\sum\limits_{l=1}^3\frac{(b_{s'}-b_{s})^2}{(\eta_l-b_s)^2(\eta_{l}-b_{s'})^2}h_l\hat{g}_{\tilde{n}}^{-1}\\
&-\sum\limits_{\tiny 1\leq s<s'\leq3}\frac{(\eta_{s}-\eta_{s'})^2\prod\limits_{\tiny1\leq l<l'\leq3}(b_{l'}-b_{l})^2}{\prod\limits_{\tiny1\leq l\leq3}(\eta_s-b_l)^2(\eta_{s'}-b_l)^2}h_sh_{s'}, \quad \tilde{n}=\{1,2,3\}\setminus\{s,s'\}.
\end{aligned}
\end{equation}
Then $\det\Omega^a=\det\tilde\Omega^a$ implies that
\begin{equation}\label{d22}
\begin{aligned}
|d_sd_{s'}|^2=\frac{\prod\limits_{l=1}^3(\eta_s-b_l)^2(\eta_{s'}-b_l)^2}{(\eta_s-\eta_{s'})^2\prod\limits_{\tiny1\leq l<l'\leq3}(b_{l'}-b_{l})^2},\\
|c_sc_{s'}|^2=\frac{\prod\limits_{l=1}^3(\eta_l-b_s)^2(\eta_{l}-b_{s'})^2}{(b_{s'}-b_{s})^2\prod\limits_{\tiny1\leq l<l'\leq3}(\eta_{l}-\eta_{l'})^2}.
\end{aligned}
\end{equation}
In addition, we also have
\begin{equation}\label{d23}
\begin{aligned}
\det\tilde\Omega=&\prod\limits_{m=1}^3\hat{h}_{m}^{-1}-\sum\limits_{j,m=1}^3\frac{g_m\hat{h}_{n_1}^{-1}\hat{h}_{n_2}^{-1}}{(\eta_l-b_m)^2}
+\sum_{3(l,m)}\frac{(\eta_l-\eta_{l'})^2(b_{m'}-b_m)^2}{\prod^*(\eta-b)}g_mg_{m'}\hat{h}_{n}^{-1}\\
&-\prod\limits_{\tiny1\leq l<l'\leq3}(\eta_{l}-\eta_{l'})^2\prod\limits_{\tiny1\leq m<m'\leq3}(b_{m'}-b_{m})^2
\prod\limits_{l,m=1}^3\frac{g_m}{(\eta_l-b_m)^2},
\end{aligned}
\end{equation}
\begin{equation}\label{d24}
\begin{aligned}
\det\Omega=&\prod\limits_{m=1}^3\hat{g}_{m}^{-1}-\sum\limits_{j,m=1}^3\frac{h_m\hat{g}_{{n}_1}^{-1}\hat{g}_{{n}_2}^{-1}}{(\eta_l-b_m)^2}
+\sum_{3(l,m)}\frac{(\eta_m-\eta_{m'})^2(b_{l'}-b_l)^2}{\prod^*(\eta-b)}h_mh_{m'}\hat{g}_{{n}}^{-1}\\
&-\prod\limits_{\tiny1\leq m<m'\leq3}(\eta_{m}-\eta_{m'})^2\prod\limits_{\tiny1\leq l<l'\leq3}(b_{l'}-b_{l})^2
\prod\limits_{l,m=1}^3\frac{h_m}{(\eta_m-b_l)^2},
\end{aligned}
\end{equation}
where the indices are defined
\begin{equation}\label{d30}
 \begin{aligned}
3(l,m):=1\leq l<l'\leq3,1\leq m<m'\leq3, \\
n=\{1,2,3\}\setminus\{l,l'\}, \quad l,l'\in\{1,2,3\}, (l<l')\\
n_1,n_2\in\{1,2,3\}\setminus\{l\}, \quad 1\leq l\leq 3, (n_1\neq n_2),
\end{aligned}
\end{equation}
and the product $\prod^*(\eta-b)$ denotes
\begin{equation}\label{d31}
 \prod^*(\eta-b)=(\eta_m-b_l)^2(\eta_m-b_{l'})^2(\eta_{m'}-b_l)^2(\eta_{m'}-b_{l'})^2.
\end{equation}
Thus $\det\Omega=-\det\tilde\Omega$ yields
\begin{equation}\label{d25}
 |c_1c_2c_3|^2=|d_1d_2d_3|^2=\frac{\prod\limits_{l,m=1}^3(\eta_m-b_l)^2}{\prod\limits_{\tiny1\leq m<m'\leq3}(\eta_{m}-\eta_{m'})^2\prod\limits_{\tiny1\leq l<l'\leq3}(b_{l'}-b_{l})^2}.
\end{equation}
From (\ref{d22}) and (\ref{d25}), we get the constraint conditions about $|c_j|$ and $|d_j|$, ($1\leq j\leq3$)
\begin{equation}\label{d26}
 |c_j|^2=\frac{\prod\limits_{l=1}^3(\eta_l-b_j)^2}{\prod\limits_{s=1,s\neq j}^3(b_s-b_j)^2}, \quad
 |d_j|^2=\frac{\prod\limits_{l=1}^3(\eta_j-b_l)^2}{\prod\limits_{s=1,s\neq j}^3(\eta_s-\eta_j)^2},
\end{equation}
and
\begin{equation}\label{d27}
\prod\limits_{\tiny1\leq m<m'\leq3}(\eta_{m}-\eta_{m'})^2=\prod\limits_{\tiny1\leq l<l'\leq3}(b_{l'}-b_{l})^2.
\end{equation}
%where $j\neq s,s', (1\leq s<s'\leq3)$.

We note that equation (\ref{d22}) is obtained from the first summation and the third summation in (\ref{d20}) and (\ref{d21}), and equation (\ref{d25}) is derived from the first product and the forth product in (\ref{d23}) and (\ref{d24}). If choose $s=n$ and $l=\tilde{n}$ in the second summation in (\ref{d20}) and (\ref{d21}), we also have some equations about $|c_s|,|d_l|$ and $\{\eta_j\},\{b_m\}$, which can also be obtained from (\ref{d26}) and (\ref{d27}), just like (\ref{d9}). It is also true for the results obtained from the second summation and the third summation in (\ref{d23}) and (\ref{d24}). Hence, for $N=3$, under the constrain condition (\ref{d26}) and (\ref{d27}), the solution of nNLS equation can be constructed, where
\begin{equation}\label{d28}
\begin{aligned}
\det\tilde\Omega^a=&\sum\limits_{\tiny 1\leq l<l'\leq3}\sum\limits_{s=1}^3\frac{(\eta_l-\eta_{l'})^2}{(\eta_l-b_s)^2(\eta_{l'}-b_s)^2}g_s\hat{h}_{n}^{-1}\\
&-\sum\limits_{\tiny 1\leq s<s'\leq3}\hat{h}_s^{-1}\hat{h}_{s'}^{-1}-\sum\limits_{\tiny 1\leq s<s'\leq3}\hat{g}_s^{-1}\hat{g}_{s'}^{-1}, \quad n=\{1,2,3\}\setminus\{l,l'\},
\end{aligned}
\end{equation}
and
\begin{equation}\label{d29}
\begin{aligned}
\det\tilde\Omega=&\prod\limits_{m=1}^3\hat{h}_{m}^{-1}-\prod\limits_{m=1}^3\hat{g}_{m}^{-1}
-\sum\limits_{j,m=1}^3\frac{g_m\hat{h}_{n_1}^{-1}\hat{h}_{n_2}^{-1}}{(\eta_l-b_m)^2}\\
&+\sum_{3(l,m)}\frac{(\eta_l-\eta_{l'})^2(b_{m'}-b_m)^2}{\prod^*(\eta-b)}g_mg_{m'}\hat{h}_{n}^{-1}.
%&-\prod\limits_{\tiny1\leq l<l'\leq3}(\eta_{l}-\eta_{l'})^2\prod\limits_{\tiny1\leq m<m'\leq3}(b_{m'}-b_{m})^2
%\prod\limits_{l,m=1}^3\frac{g_m}{(\eta_l-b_m)^2},
\end{aligned}
\end{equation}
Here the summation indices and the product are defined in (\ref{d30}) and (\ref{d31}).

It is remarked that the constraint conditions (\ref{d26}) and (\ref{d27}) for $N=\tilde{N}=3$ can be extended to general $N=\tilde{N}$, just by change 3 to $N$ in (\ref{d26}) and (\ref{d27}), which can be derived similarly from the first and last terms in (\ref{e5}) and (\ref{e5a}) by virtue of the condition (\ref{d17}). If fact, the first term of $\det\tilde\Omega^a$ takes the form
\begin{equation}\label{d33}
\begin{aligned}
-\sum\limits_{112}\prod\limits_{l,s,n_1}\frac{\hat{h}_{n_1}^{-1}g_s}{(\eta_l-b_s)^2}\prod\limits_{\tiny\begin{array}{c}
l<l'\\[-3pt]
s<s'
\end{array}}(\eta_l-\eta_{l'})^2(b_{s'}-b_s)^2=-\prod\limits_{n1}\hat{h}_{n_1}^{-1}\\
=-\sum\limits_{\tiny 1\leq s_1<\dots<s_{N-1}\leq N}\hat{h}_{s_1}^{-1}\cdots\hat{h}_{s_{N-1}}^{-1}=-\sum\limits_{j=1}^N\prod\limits_{s=1,s\neq j}^N\hat{h}_{s}^{-1},
\end{aligned}
\end{equation}
and the last term is
\begin{equation}\label{d34}
\begin{array}{c}
(-1)^N\sum\limits_{N12}\prod\limits_{l,s,n_N}\frac{\hat{h}_{n_N}^{-1}g_s}{(\eta_l-b_s)^2}
\prod\limits_{\tiny\begin{array}{c}
l<l'\\[-3pt]
s<s'
\end{array}}(\eta_l-\eta_{l'})^2(b_{s'}-b_s)^2\\
=(-1)^N\sum\limits_{j=1}^N\prod\limits_{1\leq l<l'\leq N}(\eta_l-\eta_{l'})^2\prod\limits_{\tiny\begin{array}{c}
s<s'\\[-3pt]
s,s'\neq j
\end{array}}(b_{s'}-b_s)^2\prod\limits_{l,s=1,s\neq j}^N\frac{g_s}{(\eta_l-b_s)^2}.
\end{array}
\end{equation}
%for $\sigma=N$ in $\det\tilde\Omega^a$,
%$$\hat{h}_{n_\sigma}^{-1}=0,\hat{g}_{n_\sigma}^{-1}=0$ if $\sigma=N.$$
The first term and the last term of $\det\Omega^a$ are
\begin{equation}\label{d35}
%\begin{aligned}
-\sum\limits_{112}\prod\limits_{l,s,n_1}\frac{h_s\hat{g}_{n_1}^{-1}}{(\eta_s-b_l)^2}
\prod\limits_{\tiny\begin{array}{c}
l<l'\\[-3pt]
s<s'
\end{array}}(\eta_s-\eta_{s'})^2(b_{l'}-b_l)^2=-\sum\limits_{j=1}^N\prod\limits_{s=1,s\neq j}^N\hat{g}_{s}^{-1},
%\end{aligned}
\end{equation}
\begin{equation}\label{d36}
 \begin{aligned}
(-1)^N\sum\limits_{N12}\prod\limits_{l,s,n_N}\frac{h_s\hat{g}_{n_N}^{-1}}{(\eta_s-b_l)^2}
\prod\limits_{\tiny\begin{array}{c}
l<l'\\[-3pt]
s<s'
\end{array}}(\eta_s-\eta_{s'})^2(b_{l'}-b_l)^2\\
=(-1)^N\sum\limits_{j=1}^N\prod\limits_{\tiny\begin{array}{c}
s<s'\\[-3pt]
s,s'\neq j
\end{array}}(\eta_s-\eta_{s'})^2\prod\limits_{1\leq l<l'\leq N}(b_{l'}-b_l)^2\prod\limits_{l,s=1,s\neq j}^N\frac{h_s}{(\eta_s-b_l)^2}.
 \end{aligned}
\end{equation}
By the condition $\det\tilde\Omega^a=(-1)^{N-1}\det\Omega^a$, equation (\ref{d33}) and (\ref{d36}) implies that
\begin{equation}\label{d37}
 \prod\limits_{s=1,s\neq j}^N|d_s|^2=\frac{\prod\limits_{l,s=1,s\neq j}^N(\eta_s-b_l)^2}{\prod\limits_{\tiny\begin{array}{c}
s<s'\\[-3pt]
s,s'\neq j
\end{array}}(\eta_s-\eta_{s'})^2\prod\limits_{1\leq l<l'\leq N}(b_{l'}-b_l)^2},
\end{equation}
and equations (\ref{d34}), (\ref{d35}) give
\begin{equation}\label{d38}
 \prod\limits_{s=1,s\neq j}^N|c_s|^2=\frac{\prod\limits_{l,s=1,s\neq j}^N(\eta_l-b_s)^2}{\prod\limits_{1\leq l<l'\leq N}(\eta_l-\eta_{l'})^2\prod\limits_{\tiny\begin{array}{c}
s<s'\\[-3pt]
s,s'\neq j
\end{array}}(b_{s'}-b_s)^2}.
\end{equation}
Similarly, the first term and the last term of $\det\tilde\Omega$ and $\det\Omega$ are
$$\det\tilde\Omega:\quad \prod\limits_{m=1}^N\hat{h}_m^{-1},\quad (-1)^N\prod\limits_{1\leq l,m\leq N}\frac{g_m}{(\eta_l-b_m)^2}
\prod\limits_{\tiny\begin{array}{c}
1\leq l<l'\leq N\\[-3pt]
1\leq m<m'\leq N
\end{array}}(\eta_l-\eta_{l'})^2(b_{m'}-b_m)^2;$$
$$\det\Omega:\quad \prod\limits_{m=1}^N\hat{g}_m^{-1},\quad (-1)^N\prod\limits_{1\leq l,m\leq N}\frac{h_m}{(\eta_m-b_l)^2}
\prod\limits_{\tiny\begin{array}{c}
1\leq l<l'\leq N\\[-3pt]
1\leq m<m'\leq N
\end{array}}(\eta_m-\eta_{m'})^2(b_{l'}-b_l)^2,$$
which yields
\begin{equation}\label{d39}
 \prod\limits_{s=1}^N|d_s|^2=\prod\limits_{s=1}^N|c_s|^2=\frac{\prod\limits_{l,s=1}^N(\eta_s-b_l)^2}{\prod\limits_{
1\leq s<s'\leq N}(\eta_s-\eta_{s'})^2\prod\limits_{1\leq l<l'\leq N}(b_{l'}-b_l)^2},
\end{equation}
in terms of $\det\tilde\Omega=(-1)^N\det\Omega$.
From (\ref{d37}),(\ref{d38}) and (\ref{d39}), we obtain
\begin{equation}\label{d40}
 |c_j|^2=\frac{\prod\limits_{l=1}^N(\eta_l-b_j)^2}{\prod\limits_{s=1,s\neq j}^N(b_s-b_j)^2}, \quad
 |d_j|^2=\frac{\prod\limits_{l=1}^N(\eta_j-b_l)^2}{\prod\limits_{s=1,s\neq j}^N(\eta_s-\eta_j)^2},
\end{equation}
and
\begin{equation}\label{d41}
\prod\limits_{\tiny1\leq m<m'\leq N}(\eta_{m}-\eta_{m'})^2=\prod\limits_{\tiny1\leq l<l'\leq N}(b_{l'}-b_{l})^2.
\end{equation}

\section*{Conclusions}
In this paper, we extended the Dbar-problem to discuss the cnNLS equation. In this approach, we introduced two spectral transform matrices to define two Dbar-problems. The general nonlocal relations between the cnNLS potential and the solutions of the Dbar-problems were established by Dbar-dressing method. Two sets of Dbar data were used to construct the special explicit solution of the focusing cnNLS equation. By using the Cauchy-Binet formula, We expanded the determinant solution, which was used to discuss nonlocal reductions of the cnNLS equation in detail.

%\section*{CRediT authorship contribution statement}
%Junyi Zhu: Conceptualization, Methodology, Supervision, Conducting, Visualization, Writing-Review and Editing.
%Xueru Wang: Methodology, Validation, Formal Analysis, Writing-original draft, Writing-Review and Editing.

%\section*{Declaration of interest}
%The authors declare that they have no known competing financial interests or personal relationships that could have appeared to influence the work reported in this paper.

\section*{Acknowledgments}
This work was supported by the National Natural Science Foundation of PR China [Grant number 11971442,11471295].

%\bibliography{reference_zhu}
%%\bibliographystyle{elsarticle-num}
%\bibliographystyle{elsarticle-harv}
%\bibliographystyle{unsrt}

\end{document}